\documentclass[graybox]{svmult}

\usepackage{type1cm}        
\usepackage{makeidx}         
\usepackage{graphicx}        
\usepackage{multicol}        
\usepackage[bottom]{footmisc}

\usepackage{newtxtext}       %
\usepackage[varvw]{newtxmath}       
\usepackage{subcaption}
\usepackage{algorithm, algpseudocode}
\usepackage{algcompatible}

\usepackage[acronym,shortcuts]{glossaries}
\newacronym{RPE}{RPE}{radar parameter estimation}
\newacronym{OTFS}{OTFS}{orthogonal time frequency space}
\newacronym{AFDM}{AFDM}{affine frequency division multiplexing}
\newacronym{MIMO}{MIMO}{multiple-input multiple-output}
\newacronym{SISO}{SISO}{single-input single-output}
\newacronym{ISAC}{ISAC}{integrated sensing and communications}
\newacronym{3D}{3D}{three-dimensional}
\newacronym{2D}{2D}{two-dimensional}
\newacronym{1D}{1D}{one-dimensional}
\newacronym{RX}{RX}{receiver}
\newacronym{TX}{TX}{transmitter}
\newacronym{BF}{BF}{beamforming}
\newacronym{mmWave}{mmWave}{millimeter-wave}
\newacronym{SotA}{SotA}{state-of-the-art}
\newacronym{ULA}{ULA}{uniform linear array}
\newacronym{QAM}{QAM}{quadrature amplitude modulation}
\newacronym{ISFFT}{ISFFT}{inverse symplectic finite Fourier transform}
\newacronym{SFFT}{SFFT}{symplectic finite Fourier transform}
\newacronym{AWGN}{AWGN}{additive white Gaussian noise}
\newacronym{OFDM}{OFDM}{orthogonal frequency division multiplexing}
\newacronym{OCDM}{OCDM}{orthogonal chirp division multiplexing}
\newacronym{BS}{BS}{base station}
\newacronym{UE}{UE}{user equipment}
\newacronym{DFT}{DFT}{discrete Fourier transform}
\newacronym{IDFT}{IDFT}{inverse discrete Fourier transform}
\newacronym{TD}{TD}{time-domain}
\newacronym{wlg}{wlg}{without loss of generality}
\newacronym{CP}{CP}{cyclic prefix}
\newacronym{DAFT}{DAFT}{discrete affine Fourier transform}
\newacronym{IDAFT}{IDAFT}{inverse discrete affine Fourier transform}
\newacronym{CPP}{CPP}{\textit{chirp-periodic} prefix}
\newacronym{IDZT}{IDZT}{inverse discrete Zak transform}
\newacronym{DZT}{DZT}{discrete Zak transform}
\newacronym{ICI}{ICI}{inter-carrier interference}
\newacronym{BER}{BER}{bit error rate}
\newacronym{DoF}{DoF}{degrees-of-freedom}
\newacronym{FD}{FD}{full-duplex}
\newacronym{SIMO}{SIMO}{single-input multiple-output}
\newacronym{MISO}{MISO}{multiple-input single-output}
\newacronym{AoD}{AoD}{angle-of-departure}
\newacronym{AoA}{AoA}{angle-of-arrival}
\newacronym{RF}{RF}{radio frequency}
\newacronym{SIM}{SIM}{stacked intelligent metasurfaces}
\newacronym{FPGA}{FPGA}{field programmable gate array}
\newacronym{UPA}{UPA}{uniform planar array}
\newacronym{CC}{CC}{communication-centric}
\newacronym{I/O}{I/O}{input-output}
\newacronym{iid}{i.i.d.}{independent and identically distributed}
\newacronym{IoT}{IoT}{internet of things}
\newacronym{V2X}{V2X}{vehicle-to-everything}
\newacronym{NTN}{NTN}{non-terrestrial network}
\newacronym{LEO}{LEO}{low-earth orbit}
\newacronym{THz}{THz}{terahertz}
\newacronym{EM}{EM}{electromagnetic}
\newacronym{RIS}{RIS}{reconfigurable intelligent surface}
\newacronym{DoA}{DoA}{direction-of-arrival}
\newacronym{DD}{DD}{doubly-dispersive}
\newacronym{ODDM}{ODDM}{orthogonal delay-Doppler division multiplexing}
\newacronym{LoS}{LoS}{line-of-sight}
\newacronym{NLoS}{NLoS}{non-line-of-sight}
\newacronym{6G}{6G}{sixth generation}
\newacronym{MPDD}{MPDD}{metasurfaces-parametrized DD}
\newacronym{GaBP}{GaBP}{Gaussian Belief Propagation}
\newacronym{MSE}{MSE}{mean-squared-error}
\newacronym{sIC}{soft IC}{soft interference cancellation}
\newacronym{soft RG}{soft RG}{soft replica generation}
\newacronym{BG}{BG}{belief generation}
\newacronym{SGA}{SGA}{scalar Gaussian approximation}
\newacronym{CLT}{CLT}{central limit theorem}
\newacronym{PDF}{PDF}{probability density function}
\newacronym{QPSK}{QPSK}{quadrature phase-shift keying}
\newacronym{LMMSE}{LMMSE}{linear minimum mean square error}
\newacronym{SNR}{SNR}{signal-to-noise ratio}
\newacronym{LTV}{LTV}{linear time-variant}
\newacronym{TVIRF}{TVIRF}{time-variant impulse response function}
\newacronym{FT}{FT}{Fourier transform}
\newacronym{TVTF}{TVTF}{time-variant transfer function}
\newacronym{DVIRF}{DVIRF}{Doppler-variant impulse response function}
\newacronym{CSI}{CSI}{channel state information}
\newacronym{CRLB}{CRLB}{Cram{\`e}r-Rao lower bound}
\newacronym{BCRLB}{BCRLB}{Bayesian Cram{\`e}r-Rao lower bound}
\newacronym{BBI}{BBI}{Bayesian bilinear inference}
\newacronym{ML}{ML}{maximum likelihood}
\newacronym{MUSIC}{MUSIC}{multiple signal classification}
\newacronym{MU}{MU}{multi-user}
\newacronym{ROOT-MUSIC}{ROOT-MUSIC}{ROOT multiple signal classification}
\newacronym{JCAS}{JCAS}{joint communication and sensing}
\newacronym{JCR}{JCR}{joint communications and radar}
\newacronym{ROI}{ROI}{region of interest}
\newacronym{MF}{MF}{matched-filter}
\newacronym{ISI}{ISI}{inter-symbol interference}
\newacronym{RMSE}{RMSE}{root mean square error}
\newacronym{ESPRIT}{ESPRIT}{estimation of signal parameters via rotational invariant techniques}
\newacronym{JCDE}{JCDE}{joint channel and data estimation}
\newacronym{PDA}{PDA}{probabilistic data association}
\newacronym{PMF}{PMF}{probability mass function}
\newacronym{PBiGaBP}{PBiGaBP}{parametric bilinear Gaussian belief propagation}
\newacronym{PBiGAMP}{PBiGAMP}{parametric bilinear generalized approximate message passing}
\newacronym{GAMP}{GAMP}{generalized approximate message passing}
\newacronym{EP}{EP}{expectation propagation}
\newacronym{DAF}{DAF}{discrete affine Fourier}
\newacronym{P/S}{P/S}{parallel-to-serial}
\newacronym{S/P}{S/P}{serial-to-parallel}
\newacronym{SBL}{SBL}{sparse Bayesian learning}
\newacronym{MPA}{MPA}{message passing algorithms}
\newacronym{VGA}{VGA}{vector Gaussian approximation}
\newacronym{SIC}{SIC}{self-interference cancellation}
\newacronym{NMSE}{NMSE}{normalized mean square error}
\newacronym{KL}{KL}{Kullback-Leibler}
\newacronym{i.i.d.}{i.i.d.}{independent and identically distributed}

\newcommand{\trans}[0]{^{\mathsf{T}}}
\newcommand{\herm}[0]{^{\mathsf{H}}}
\newcommand{\vect}[1]{\mathrm{vec}\!\left({#1}\right)}
\newcommand{\Real}[1]{\Re\{{#1}\}}
\newcommand{\Imag}[1]{\Im\{{#1}\}}

\makeindex             

\begin{document}

\title*{Metasurfaces-Integrated Doubly-Dispersive MIMO: Channel Modeling and Optimization}
\author{Kuranage Roche Rayan Ranasinghe\orcidID{0000-0002-6834-8877},\\ 
Hyeon Seok Rou\orcidID{0000-0003-3483-7629}, \\
Iv{\'a}n Alexander Morales Sandoval\orcidID{0000-0002-8601-5451}, \\
Giuseppe Thadeu Freitas de Abreu\orcidID{0000-0002-5018-8174} and \\
George C. Alexandropoulos\orcidID{0000-0002-6587-1371}
}
\institute{Kuranage Roche Rayan Ranasinghe \at Constructor University (previously known as Jacobs University Bremen), Campus Ring 1, 28759 Bremen, Germany, \email{kranasinghe@constructor.university}
\and 
Hyeon Seok Rou \at Constructor University (previously known as Jacobs University Bremen), Campus Ring 1, 28759 Bremen, Germany, \email{hrou@constructor.university}
\and
Iv{\'a}n Alexander Morales Sandoval \at Constructor University (previously known as Jacobs University Bremen), Campus Ring 1, 28759 Bremen, Germany, \email{imorales@constructor.university}
\and
Giuseppe Thadeu Freitas de Abreu \at Constructor University (previously known as Jacobs University Bremen), Campus Ring 1, 28759 Bremen, Germany, \email{gabreu@constructor.university}
\and
George C. Alexandropoulos \at National and Kapodistrian University of Athens, 16122 Athens, Greece, \email{alexandg@di.uoa.gr}
}
%
%
\titlerunning{Metasurfaces-Integrated Doubly-Dispersive MIMO: Modeling and Optimization}  
\authorrunning{Ranasinghe et al.}
\maketitle

\abstract{   
The \ac{DD} channel structure has played a pivotal role in wireless communications, particularly in high-mobility scenarios and \ac{ISAC}, due to its ability to capture the key fading effects experienced by a transmitted signal as it propagates through a dynamic medium. 
However, extending the \ac{DD} framework to \ac{MIMO} systems, especially in environments artificially enhanced by \acp{RIS} and \ac{SIM}, remains a challenging open problem.
In this chapter, a novel \ac{MPDD} channel model that integrates an arbitrary number of \acp{RIS}, while also incorporating \ac{SIM} at both the transmitter and receiver is introduced. 
Next, the application of this model to some key waveforms optimized for \ac{DD} environments -- namely \ac{OFDM}, \ac{OTFS}, and \ac{AFDM} -- is discussed.
Finally, the programmability of the proposed model is highlighted through an illustrative application, demonstrating its potential for enhancing waveform performance in \ac{SIM}-assisted wireless systems. 
}

\glsresetall
\newpage
\section{Introduction}
\label{sec:Introduction}

Next generation wireless communication systems are anticipated to enable a diverse range of new applications, including \ac{V2X} and aerial communications \cite{ChenCSM2017}, \ac{IoT} networks \cite{NguyenIoTJ2022}, and \acp{NTN} such as \ac{LEO} satellite networks \cite{ShiNetwork2024}.
Due to the intrinsic nature of the resulting high-mobility environment in these systems \cite{LiCOMMST2022}, there is a great demand for accurate channel models and robust detectors/estimators to encapsulate such surroundings.

However, high-mobility communication scenarios pose significant challenges due to the time-frequency selectivity in time-varying multipath conditions \cite{WangTWC2006}. 
This necessitates the use of \ac{DD} channel models \cite{LiuTIT2004, Bliss_Govindasamy_2013,Tri_polarized, Rou_SPM_2024}, which have recently been leveraged for devising various advanced communication techniques \cite{SurabhiTWC2019, BomfinTWC2021, PfadlerTWC2024, LiangTWC2024, HaifTVT2024, ZhangJSAC2024}.
A key development in this area is the application of \ac{DD} channel models for \ac{RPE}, leading to \ac{CC} \ac{ISAC} \cite{position_aided,HyeonTWC2024, HyeonAsilomar2022, GaudioTWC2020, Mohammed_BITS_2022,ISAC_FD, ISAC_FD_MIMO,Gupta_OJCS_2024, Ranasinghe_ICASSP_2024, KuranageTWC2024,6gdisac,ISAC_exponent,ISAC_XL_MIMO} which enables both sensing and communications functionalities. \ac{CC}-\ac{ISAC} offers a compelling \ac{ISAC} solution in high-mobility scenarios by integrating \ac{RPE} techniques with traditional signal processing methods, such as de-chirping in \ac{AFDM} \cite{Bemani_WCL_2024}, bilinear inference \cite{Parker_TSP_2014, IimoriTWC2021, TakahashiTWC2023}, and blind covariance-based detection \cite{YangCommL2011, BaoACCESS2019}. 
This allows \ac{ISAC} to be implemented in mono-, bi-, and multi-static configurations \cite{RanasingheWCNC2024}.

The effective waveform design constitutes a crucial aspect for the use of \ac{DD} channels. 
While many such waveforms exist in literature, three of the primary candidates include:
\begin{itemize}
\item \textbf{\Ac{OFDM}}: A widely adopted waveform with useful \ac{RPE} properties \cite{LiyanaarachchiTWC2024}, but susceptible to severe inter-carrier interference in high-mobility scenarios \cite{Gaudio_TWC_2022}.
\item \textbf{\Ac{OTFS}}: A \ac{2D} modulation scheme embedding data in the delay-Doppler domain, offering robustness to \ac{DD} effects, but requiring higher implementation complexity \cite{SurabhiTWC2019} relative to \ac{OFDM}.
\item \textbf{\Acf{AFDM}}: A recently proposed alternative \cite{Ni_ISWCS_2022, Bemani_TWC_2023, RouAsilimoar2024} achieving optimal diversity in \ac{DD} channels with a similar implementation complexity to \ac{OFDM}, though facing practical challenges in chirp signal generation \cite{Srivastava_ISES2018}.
\end{itemize}

Recent studies have shown that these waveforms share structural similarities within the \ac{DD} model, enabling unified system design approaches \cite{Rou_SPM_2024}. 
However, existing unifying work is limited to \ac{SISO} systems, despite the established role of \ac{MIMO} in modern wireless networks, including emerging paradigms such as extremely large \ac{MIMO} \cite{XLMIMO_tutorial, NF_beam_tracking} and reconfigurable metasurfaces \cite{Tsinghua_RIS_Tutorial, BAL2024}.

Motivated by these advancements, in this chaper, a novel \ac{MPDD} \ac{MIMO} channel model incorporating multi-functional metasurfaces, such as \acp{RIS} \cite{LiuCommST2021, RIS_influence,Tsinghua_RIS_Tutorial, BAL2024}, dynamic metasurfaces antennas~\cite{DMAs_1,DMAs_2}, and \ac{SIM} \cite{PG13, ZKW+18, HLC+19} is proposed, which is amenable to optimization for enhancing high-mobility wireless communications. 
The key contributions of this chapter are the following:
\begin{itemize}
\item Development of a \ac{MPDD} \ac{MIMO} channel model integrating \ac{TX} and \ac{RX} \ac{SIM} alongside multiple \acp{RIS}, suitable for high-mobility and \ac{CC}-\ac{ISAC} applications.
\item Derivation of end-to-end \ac{TD} received signal expressions and effective channel matrices for \ac{OFDM}, \ac{OTFS}, and \ac{AFDM}, facilitating \ac{ISAC} objectives.
\item Optimization of \ac{SIM} for enhanced signal reception, employing a \ac{GaBP}-based detector, demonstrating significant performance gains over conventional \ac{DD} models.
\end{itemize}

\textit{Notation:} All scalars are represented by upper or lowercase letters, while column vectors and matrices are denoted by bold lowercase and uppercase letters, respectively.
The diagonal matrix constructed from vector $\vec{a}$ is denoted by diag($\vec{a}$), while $\vec{A}\trans$, $\vec{A}\herm$, $\vec{A}^{1/2}$, and $[\vec{A}]_{i,j}$ denote the transpose, Hermitian, square root, and the $(i,j)$-th element of a matrix $\vec{A}$, respectively.
The convolution and Kronecker product are respectively denoted by $*$ and $\otimes$, while $\vec{I}_N$ and $\vec{F}_N$ represent the $N\times N$ identity and the normalized $N$-point \ac{DFT} matrices, respectively.
The sinc function is expressed as $\text{sinc}(a) \triangleq \frac{\sin(\pi a)}{\pi a}$, whereas $\jmath\triangleq\sqrt{-1}$ denotes the elementary complex number.

\section{Preliminaries}
\label{sec:Preliminaries}

In this section, some preliminary background on the \ac{SISO} and \ac{MIMO} \ac{DD} channel model as well as the \ac{RIS} and \ac{SIM} technologies is provided.

\subsection{SISO Doubly-Dispersive Channel Model}
\label{subsec:SISO_DD_Channel_Model}

Consider a wireless channel between a single-antenna transmitter and single-antenna receiver, modeled using $P$ significant propagation paths. 
Each $p$-th resolvable path, with $p \in {1, \dots, P}$, is characterized by a complex fading coefficient $h_p \in \mathbb{C}$, a path delay $\tau_p \in [0, \tau^\mathrm{max}]$, and a Doppler shift $\nu_p \in [-\nu^\mathrm{max}, +\nu^\mathrm{max}]$.
The delay and Doppler spreads of this \ac{DD} channel are determined by the maximum delay $\tau^\mathrm{max}$ in seconds and the maximum Doppler shift $\pm \nu^\mathrm{max}$ in Hz. 

The channel behavior is described by the \ac{LTV}\footnote{The term ``\acf{LTV} system'' is distinct from linear systems with time-varying delays (i.e., $\tau_p(t)$), commonly also referred to as \ac{LTV}. Herein, only time-invariant delays are considered, in line with prior studies \cite{Bliss_13, Wei_WC21,Bemani_TWC_2023}.} relationship between the input and output signals, and is most commonly represented as a \ac{TVIRF} in the time-delay domain given by \cite{Rou_SPM_2024}
\begin{equation}
h^\text{TD}(t, \tau) \triangleq \sum_{p=1}^{P} h_p  e^{\jmath 2\pi \nu_p t} \delta(\tau - \tau_p),
\label{eq:SISO_DD_channel_model}
\end{equation}
where $t$ and $\tau$ denote instantaneous time and delay, respectively, and $\delta(x)$ is the unit impulse function. Alternative representations also exist in other domains via appropriate linear transforms \cite{Healy_LCT15}. 
For example, the time-frequency representation, known as \ac{TVTF}, is obtained through a \ac{FT} over the delay domain, $\mathcal{F}[\cdot]$, and is represented as follows:
\begin{equation}
\label{eq:SISO_DD_channel_model_TVTF}
h^{\text{TF}}(t, f) \triangleq 
\mathcal{F}\left[ h^{\text{TD}}(t, \tau)\right]
= \sum_{p=1}^{P} h_p e^{\jmath 2\pi \nu_p t} e^{-\jmath 2\pi \tau_p f}.
\end{equation}
Similarly, the delay-Doppler representation, termed \ac{DVIRF}, is obtained via an \ac{FT} over the time domain as
\begin{equation}
    \label{eq:SISO_DD_channel_model_DVIRF}
h^{\text{DD}}(\nu, \tau) \triangleq \sum_{p=1}^{P} h_p  \delta(\nu - \nu_p) \delta(\tau - \tau_p).
\end{equation}

\begin{remark}[\ac{DD} Channel Assumptions]
 The channel models presented in expressions \eqref{eq:SISO_DD_channel_model}-- \eqref{eq:SISO_DD_channel_model_DVIRF} are based on a practical approximation of the true \ac{DD} wireless channel (resulting from the incompatibility of the bandwidth-limited assumption and the temporally-limited assumption of the signals), represented in terms of a finite number of significant delay and Doppler frequency taps, which is known to generally work well \cite{Hong_2022, Bliss_13}, especially under the underspread environment assumption, i.e., the maximum delay spread $\tau_\mathrm{max} - \tau_\mathrm{min}$ is smaller than $T$, the maximum Doppler spread $\nu_\mathrm{max} - \nu_\mathrm{min}$ is smaller than $\frac{1}{T}$, and $\tau_\mathrm{max} \nu_\mathrm{max} <\!< 1$, where $T$ is the finite signal period measured in seconds.
\end{remark}

Hereafter, \ac{wlg}, the time-delay representation in expression~\eqref{eq:SISO_DD_channel_model} is chosen to simplify the upcoming derivations for arbitrary \ac{TD} signals.
For receiver designs based on the \ac{SISO} \ac{DD} channel model leveraged for \ac{ISAC}, the reader is referred to \cite{KuranageTWC2024}.

\subsection{MIMO Doubly-Dispersive Channel Model}
\label{subsec:MIMO_DD_Channel_Model}

Let $\phi\in [0,\pi]$ denote the arbitrary \ac{AoA} or \ac{AoD} of a channel propagation path to (or from) a \ac{ULA} with $A$ antenna elements.
Then, the array response vector $\vec{a} (\phi) \in \mathbb{C}^{A \times 1}$ is defined as
\begin{equation}
\label{eq:ULA_response_vector_transmit}
\vec{a} (\phi) \triangleq \tfrac{1}{\sqrt{A}} \Big[ 1, e^{-\jmath\frac{2\pi}{\lambda}d \sin(\phi)}, \dots, e^{-\jmath\frac{2\pi}{\lambda}(A-1)d \sin(\phi)} \Big]\trans,
\end{equation}
where $\lambda$ indicates the wavelength and $d$ is the antenna spacing, which is usually set as $d=\lambda/2$ \cite{SrivaTWC2022}.

Similarly, for a \ac{UPA} with $B \triangleq B_x B_z$ elements\footnote{Without loss of generality, the \ac{UPA} is aligned parallel to the $y$ direction with elements occupying space in the $x$ and $z$ dimensions. The generalization to arbitrary axes is trivial; some other orientations are discussed in~\cite{AnJSAC2024,NF_beam_tracking}.
}, the response vector corresponding to a path impinging onto (or outgoing from) the array at the elevation and azimuth angles $\theta\in [0,\pi]$ and $\phi \in [-\frac{\pi}{2},\frac{\pi}{2}]$ is given by \cite{AnJSAC2023}
\begin{equation}
\label{eq:SIM_transmit_UPA_response vector}
\vec{b}(\phi,\theta) \triangleq \tfrac{1}{\sqrt{B_x B_z}}  \vec{b}_x(\phi,\theta) \otimes \vec{b}_z(\theta) \in \mathbb{C}^{B \times 1},
\end{equation}
where the $x$- and $z$-axis steering vectors $\vec{b}_x(\phi,\theta) \in \mathbb{C}^{B_x \times 1}$ and $\vec{b}_z(\theta) \in \mathbb{C}^{B_z \times 1}$ are respectively defined as
\begin{subequations}
\vspace{-1ex}
\begin{equation}
\label{eq:steering_x_axis} 
\vec{b}_x(\phi,\theta) \triangleq \Big[1, e^{-\jmath \frac{2\pi d_x}{\lambda} \sin(\phi) \sin(\theta)}, \dots, e^{-\jmath \frac{2\pi d_x}{\lambda} (B_x - 1)  \sin(\phi) \sin(\theta)} \Big]\trans,
\end{equation}
and 
\begin{equation}
\label{eq:steering_z_axis} 
\!\!\vec{b}_z(\theta) \triangleq \!\Big[1, e^{-\jmath \frac{2\pi d_z}{\lambda} \cos(\theta)}, \dots, e^{-\jmath \frac{2\pi d_z}{\lambda} (B_z - 1) \cos(\theta)} \Big]\trans,
\end{equation}
\end{subequations}
with $d_x$ and $d_z$ being the element spacing in the \ac{UPA}'s $x$- and $z$-axis directions, respectively, which are usually set as $d_x = d_z = \lambda/2$.

Subsequently, the \ac{MIMO} equivalent of expression~\eqref{eq:SISO_DD_channel_model} defined in Section~\ref{subsec:SISO_DD_Channel_Model}, for the case with $N_\mathrm{T}$ transmit antennas and $N_\mathrm{R}$ receive antenna,s $\vec{H}^\text{TD}(t, \tau) \in \mathbb{C}^{N_\mathrm{R} \times N_\mathrm{T}}$, is given by
\begin{equation}
    \vec{H}^\text{TD}(t, \tau) \triangleq \sqrt{\tfrac{N_\mathrm{T} N_\mathrm{R}}{P}} \sum_{p=1}^{P} h_p \cdot e^{\jmath 2\pi \nu_p t}  \delta(\tau - \tau_p) \vec{a}_\mathrm{R} (\phi_p^{\rm in}) \vec{a}_\mathrm{T}\herm (\phi_p^{\rm out}),
\label{eq:MIMO_DD_channel_model}
\end{equation}
where $\vec{a}_\mathrm{R} (\phi_p^{\rm in})$ and $\vec{a}_\mathrm{T} (\phi_p^{\rm out})$ are the receive and transmit array response vectors, respectively, corresponding to the $p$-th path with \ac{AoA} $\phi_p^{\rm in}$ and \ac{AoD} $\phi_p^{\rm out}$, respectively.

\subsection{RIS Response Model}
\label{subsec:RIS_Model}

Consider a wireless scenario involving $K$ \acp{RIS}, where each \ac{RIS} comprises $J \triangleq J_x J_z$ reflective meta-atoms with tunable responses. Here, $J_x$ and $J_z$ represent the number of meta-atoms along the $x$- and $z$-axes, respectively. The reflective phase response of each $j$-th meta-atom ($j \in \{1,\ldots,J\}$) within the $k$-th \ac{RIS} ($k \in \{1,\ldots,K\}$) is denoted by $\phi^{k}_{j} \in [0,2\pi)$. 
Consequently, the phase configuration of the $k$-th \ac{RIS} can be represented by the $J \times J$ phase matrix $\vec{\Phi}_k$, which is given by~\cite{Marconi}:
\begin{equation}
\vec{\Phi}_k  \triangleq \text{diag}\bigg( \!\Big[ e^{\jmath\phi^{k}_{1}}, \dots, e^{\jmath\phi^{k}_{J}} \Big]\! \bigg).
\label{eq:diagona_shift_matrix_RIS}
\end{equation}

\subsection{SIM Response Model}
\label{subsec:SIM_Model}

Consider a \ac{SIM} consisting of $Q$ layers of transmissive metasurfaces arranged in parallel with minimal separation. Each metasurface is composed of $M \triangleq M_x M_z$ response-tunable meta-atoms, 
where $M_x$ and $M_z$ denote the number of meta-atoms along the $x$- and $z$-axis per layer, respectively. 
Next, the following $M \times M$ matrix is defined, which encapsulates the effective tunable phase shifts of all $M$ meta-atoms embedded in each $q$-th ($q \in \mathcal{Q} \triangleq \{1, \dots, Q\}$) metasurface layer of the \ac{SIM}, as follows:
\begin{equation}
\vec{\Psi}_q  \triangleq \text{diag}\bigg( \!\Big[ e^{\jmath\zeta^{q}_{1}}, \dots, e^{\jmath\zeta^{q}_{M}} \Big]\! \bigg),
\label{eq:diagona_shift_matrix_per_layer}
\end{equation}
where $\zeta^{q}_{m} \in [0,2\pi)$ $\forall q \in \mathcal{Q}$ and $\forall m \in \mathcal{M} \triangleq \{ 1, \dots, M \}$ represents the transmissive phase response of the $m$-th meta-atom lying on the $q$-th metasurface layer. 

The transmission matrix between each $(q-1)$-th and $q$-th layer of the \ac{SIM}, for all $q \in \mathcal{Q} \backslash { 1}$, is denoted as $\vec{\Gamma}_q \in \mathbb{C}^{M \times M}$.
According to Rayleigh-Sommerfeld diffraction theory, each $(m,m')$-th element (where $m,m' \in \mathcal{M}$) of $\vec{\Gamma}_q$ represents the diffraction coefficient between the $m'$-th meta-atom on the $(q-1)$-th metasurface and the $m$-th meta-atom on the $q$-th metasurface, and is given by \cite{AnJSAC2023}:
\begin{equation}
\label{eq:diffraction_coeff}
\gamma^{q}_{m,m'} \triangleq \frac{\rho_t \cos\big(\epsilon^{q}_{m,m'}\big)}{\mathrm{d}^{q}_{m,m'}}  \bigg( \frac{1}{2\pi  \mathrm{d}^{q}_{m,m'}} - \frac{\jmath}{\lambda} \bigg)  e^{\jmath2\pi \frac{\mathrm{d}^{q}_{m,m'}}{\lambda}},
\end{equation}
where $\rho_t$ is the square measure occupied by each meta-atom in the \ac{SIM}, $\epsilon^{q}_{m,m'}$ denotes the angle between the propagation and normal direction of the $(q-1)$-th metasurface layer, and $\mathrm{d}^{q}_{m,m'}$ represents the propagation distance. 

Placing the \ac{SIM} very close to an $N_\mathrm{T}$-element TX \ac{ULA} with adjacent inter-element spacing $\lambda / 2$, let us define $\vec{\Gamma}_{1} \triangleq [\vec{\gamma}^1_1, \dots, \vec{\gamma}^1_{N_\mathrm{T}}] \in \mathbb{C}^{M \times N_\mathrm{T}}$ with $\vec{\gamma}^1_{n_\mathrm{T}} \in \mathbb{C}^{M \times 1}$ (with $n_\mathrm{T}=1,\ldots,N_\mathrm{T}$) representing the transmission vector from the $n_\mathrm{T}$-th transmit antenna to the innermost metasurface layer of the \ac{SIM}, whose $m$-th element $\gamma^1_{m,n_\mathrm{T}}$ is obtained by substituting $\epsilon^{q}_{m,m'}$ and $\mathrm{d}^{q}_{m,m'}$ in expression~\eqref{eq:diffraction_coeff} with $\epsilon^1_{m,n_\mathrm{T}}$ and $\mathrm{d}^1_{m,n_\mathrm{T}}$, respectively. 

Bridging the concepts introduced above and defining $\vec{\mathcal{Z}}\triangleq\{\vec{\Psi}_1,\ldots,\vec{\Psi}_Q\}$, the overall $M\times N_\mathrm{T}$ propagation matrix from the \ac{TX} antenna elements to the meta-atoms of the $Q$-th \ac{SIM} layer can be expressed as follows:
\begin{equation}
\vec{\Upsilon}_\mathrm{T}(\vec{\mathcal{Z}}) \triangleq \prod_{q=1}^Q \vec{\Psi}_{Q-q+1} \vec{\Gamma}_{Q-q+1}.
\label{eq:transmit_SIM_full}
\end{equation}

Similarly, a \ac{SIM} of $\tilde{\mathcal{Q}}$ layers of transmissive metasurfaces, each comprising $\tilde{M} \triangleq \tilde{M}_x \tilde{M}_z$ response-tunable meta-atoms with $\tilde{M}_x$ and $\tilde{M}_z$ being the number of meta-atoms in the $x$- and $z$-axis on each layer, respectively, placed very close to an $N_\mathrm{R}$-element RX \ac{ULA} with adjacent inter-element spacing $\lambda / 2$, results in the following overall $N_\mathrm{R}\times \tilde{M}$ propagation matrix from the SIM to the RX antenna array:
\begin{equation}
\vec{\Upsilon}_\mathrm{R}(\tilde{\vec{\mathcal{Z}}}) \triangleq \prod_{\tilde{q}=1}^{\tilde{Q}} \vec{\Xi}_{\tilde{q}} \vec{\Delta}_{\tilde{q}},
\label{eq:receive_SIM_full}
\end{equation}
where $\vec{\Xi}_{\tilde{q}} \in \mathbb{C}^{\tilde{M} \times \tilde{M}}$ $\forall \tilde{q} \in\{2,\ldots,\tilde{Q}\}$ is the transmission matrix between the $\tilde{q}$-th and $(\tilde{q}-1)$-th layer of the \ac{SIM}, whose elements are defined similar to~\eqref{eq:diffraction_coeff} as
\begin{equation}
\label{eq:diffraction_coeff_rx}
\xi^{\tilde{q}}_{\tilde{m},\tilde{m}'} \triangleq \frac{\rho_r \cos\big(\tilde{\epsilon}^{\tilde{q}}_{\tilde{m},\tilde{m}'}\big)}{\tilde{\mathrm{d}}^{\tilde{q}}_{\tilde{m},\tilde{m}'}}  \bigg( \frac{1}{2\pi  \tilde{\mathrm{d}}^{\tilde{q}}_{\tilde{m},\tilde{m}'}} - \frac{\jmath}{\lambda} \bigg)  e^{\jmath2\pi \frac{\tilde{\mathrm{d}}^{\tilde{q}}_{\tilde{m},\tilde{m}'}}{\lambda}}.
\end{equation}
In this expression, $\rho_r$ denotes the square measure occupied by each meta-atom in the \ac{RX}-\ac{SIM}, $\tilde{\epsilon}^{\tilde{q}}_{\tilde{m},\tilde{m}'}$ is the angle between the propagation and normal direction of the $(\tilde{q}-1)$-th metasurface layer, and $\tilde{\mathrm{d}}^{\tilde{q}}_{\tilde{m},\tilde{m}'}$ corresponds to the propagation distance. Subsequently, $\vec{\Xi}_{1} \triangleq [\vec{\xi}^1_1, \dots, \vec{\xi}^1_{N_\mathrm{R}}]\trans \in \mathbb{C}^{N_\mathrm{R} \times \tilde{M}}$ with $\vec{\xi}^1_{n_\mathrm{R}} \in \mathbb{C}^{\tilde{M} \times 1}$ ($n_\mathrm{R}=1,\ldots,N_\mathrm{R}$) denoting the transmission vector from the $n_\mathrm{R}$-th receive antenna to the innermost \ac{SIM} layer, whose $\tilde{m}$-th element $\xi^1_{\tilde{m},n_\mathrm{R}}$ with $m \in \tilde{\mathcal{M}}\triangleq\{ 1, \dots, \tilde{M} \}$ is defined similarly to $\gamma^1_{m,n_\mathrm{T}}$, but using expression~ \eqref{eq:diffraction_coeff_rx}. 
Finally, $\vec{\Delta}_{\tilde{q}}$ $\forall\tilde{q} \in \tilde{\mathcal{Q}}\triangleq\{1,\ldots,\tilde{Q}\}$ defined similar to expression~ \eqref{eq:diagona_shift_matrix_per_layer} including the effective tunable phase responses 
$\tilde{\zeta}^{\tilde{q}}_{\tilde{m}}\in [0,2\pi)$ $\forall \tilde{q} \in \tilde{\mathcal{Q}}$ and $\forall \tilde{m} \in \tilde{\mathcal{M}}$ with $\tilde{\vec{\mathcal{Z}}}\triangleq\{\vec{\Delta}_1,\ldots,\vec{\Delta}_{\tilde{Q}}\}$ is given by
\begin{equation}
\vec{\Delta}_{\tilde{q}}  \triangleq \text{diag}\bigg( \!\Big[ e^{\jmath \tilde{\zeta}^{\tilde{q}}_{1}}, \dots, e^{\jmath \tilde{\zeta}^{\tilde{q}}_{\tilde{M}}} \Big]\! \bigg).
\label{eq:diagona_shift_matrix_per_layer_rx}
\end{equation}

\section{The Proposed MPDD MIMO Channel Model}
\label{sec:MPDD_MIMO_Channel_Model}

\begin{figure}[t]
    \includegraphics[scale=.34]{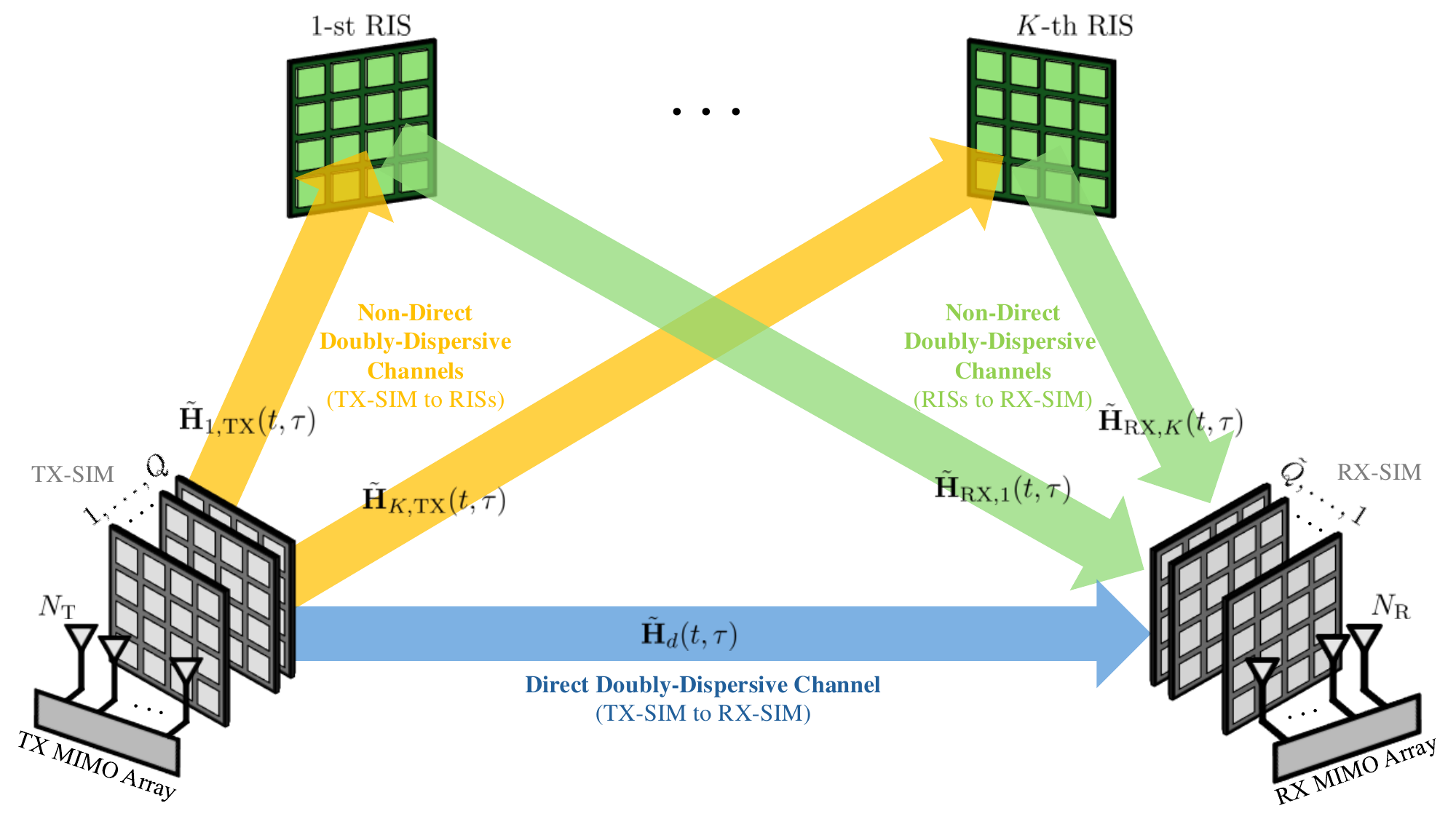}
    \caption{The considered MPDD \ac{MIMO} system for high-mobility scenarios, which includes two \ac{SIM}, one placed very close to the \ac{TX} and the other very close to the \ac{RX}, and $K$ \acp{RIS} within the wireless propagation environment of interest.}
    \label{fig:system_model}     
\end{figure}

Consider the \ac{MIMO} system illustrated in Fig.~\ref{fig:system_model}, where a transmitter with an $N_\mathrm{T}$-element \ac{ULA} and a $Q$-layered \ac{SIM}, and a receiver equipped with a $\tilde{Q}$-layered \ac{SIM} and a \ac{ULA}-based front-end with $N_\mathrm{R}$ antennas, establish a communications link.
Both the \ac{TX} and \ac{RX} \ac{SIM} are placed very close to their respective antenna arrays, as described in Section~\ref{subsec:SIM_Model}, and are respectively denoted as \ac{TX}-\ac{SIM} and \ac{RX}-\ac{SIM}.
In addition, as clarified in the figure, the \ac{MIMO} system operates within a smart wireless environment~\cite{RIS_influence} comprising $K$ \acp{RIS} each consisting of $J$ response-tunable reflective meta-atoms, as described in Section~\ref{subsec:RIS_Model}.

The complex-valued $N_\mathrm{R} \times N_\mathrm{T}$ end-to-end \ac{MPDD} smart wireless \ac{MIMO} channel, which is parametrized via the configurations of the \ac{TX} and \ac{RX} \ac{SIM} (i.e., $\vec{\mathcal{Z}}$ and $\tilde{\vec{\mathcal{Z}}}$, respectively) as well as the \acp{RIS} (i.e., $\vec{\mathcal{F}}\triangleq\{\vec{\Phi}_1,\ldots,\vec{\Phi}_K\}$), can be expressed as follows\footnote{The proposed MPDD \ac{MIMO} channel model holds also for cases where any of the \ac{SIM} or \acp{RIS} has a non-local structure \cite{NeriniTWC2024,WMA2024,BD_RIS_Opt}.}:
\begin{equation}
\label{eq:prop_SIM_channel}
\!\!\!\!\vec{H}(\vec{\mathcal{Z}},\tilde{\vec{\mathcal{Z}}},\vec{\mathcal{F}},t,\tau) \!\triangleq\! \vec{\Upsilon}_\mathrm{R}(\tilde{\vec{\mathcal{Z}}})\vec{R}_\mathrm{RX}^{1/2} \tilde{\vec{H}}(\vec{\mathcal{F}},t,\tau) \vec{R}_\mathrm{TX}^{1/2}\vec{\Upsilon}_\mathrm{T}(\vec{\mathcal{Z}}),\!\!
\end{equation}    
where $\vec{R}_\mathrm{TX} \in \mathbb{C}^{M \times M}$ and $\vec{R}_\mathrm{RX} \in \mathbb{C}^{\tilde{M} \times \tilde{M}}$ are, respectively, the spatial correlation matrices at the outermost -- $i.e.$, respectively the $\mathcal{Q}$-th and $\tilde{\mathcal{Q}}$-th -- layer of the \ac{TX}-\ac{SIM} and \ac{RX}-\ac{SIM}, due to the sub-wavelength spacing of adjacent meta-atoms~\cite{corr_1,corr_2,corr_3}.

Following \cite{AnJSAC2023}, each $(m,m')$-th and $(\tilde{m},\tilde{m}')$-th elements of $\vec{R}_\mathrm{TX}$ and $\vec{R}_\mathrm{RX}$ in expression~\eqref{eq:prop_SIM_channel} are respectively defined as $[\vec{R}_\mathrm{TX}]_{m,m'} \triangleq \text{sinc}\left(2\mathrm{d}_{m,m'}/\lambda\right)$ and $[\vec{R}_\mathrm{RX}]_{\tilde{m},\tilde{m}'} \triangleq \text{sinc}\big(2\tilde{\mathrm{d}}_{\tilde{m},\tilde{m}'}/\lambda\big)$, while the complex-valued $\tilde{M} \times M$ \ac{DD} \acp{RIS}-parametrized channel matrix $\tilde{\vec{H}}(\vec{\mathcal{F}},t,\tau)$ is given by
\begin{equation}
\label{eq:MIMO_TD_channel_general_SIM}
\tilde{\vec{H}}(\vec{\mathcal{F}},t,\tau) \!\triangleq\! \tilde{\vec{H}}_d(t,\tau) +\! \sum_{k=1}^{K} \tilde{\vec{H}}_{\mathrm{RX},k}(t,\tau) \vec{\Phi}_k \tilde{\vec{H}}_{k,\mathrm{TX}}(t,\tau),\!
\end{equation}
where $\tilde{\vec{H}}_d(t,\tau) \in \mathbb{C}^{\tilde{M} \times M}$ represents the direct $P$-path \ac{DD} \ac{MIMO} channel between the $\tilde{M}$-element $\tilde{\mathcal{Q}}$-th layer of the \ac{RX}-\ac{SIM} and the $M$-element $\mathcal{Q}$-th layer of the \ac{TX}-\ac{SIM}, which is defined as follows:
\begin{equation}
\label{eq:MIMO_TD_channel_general_SIM_LOS}
\tilde{\vec{H}}_d(t,\tau) \triangleq \sqrt{\tfrac{M\tilde{M}}{P}} \sum_{p=1}^P h_p e^{\jmath 2\pi \nu_p t} \delta\left(\tau-\tau_p\right) \vec{b}_\mathrm{R}\left(\phi_p^{\rm in},\theta_p^{\rm in}\right) \vec{b}_\mathrm{T}\herm\left(\phi_p^{\rm out},\theta_p^{\rm out}\right),
\end{equation}
where $\vec{b}_\mathrm{T}(\cdot,\cdot) \in \mathbb{C}^{M \times 1}$ and $\vec{b}_\mathrm{R}(\cdot,\cdot) \in \mathbb{C}^{\tilde{M} \times 1}$ are respectively the \ac{UPA} response vectors for the \ac{TX}-\ac{SIM} and \ac{RX}-\ac{SIM} defined in expression~\eqref{eq:SIM_transmit_UPA_response vector} in Section~\ref{subsec:MIMO_DD_Channel_Model}, with $(\phi_p^{\rm in},\theta_p^{\rm in})$ and $(\phi_p^{\rm out},\theta_p^{\rm out})$ being the pairs of azimuth and elevation \acp{AoA} and \acp{AoD}, respectively, for each $p$-th signal propagation path. The \ac{UPA} response matrix associated with a given $p$-th path is defined as:
\begin{align}
\label{eq:Bmatrix}
\vec{B}_p &\triangleq  \vec{b}_\mathrm{R}\left(\phi_p^{\rm in},\theta_p^{\rm in}\right) \vec{b}_\mathrm{T}\herm\left(\phi_p^{\rm out},\theta_p^{\rm out}\right) .
\end{align}

Finally, for the sake of clarity, the matrices $\tilde{\vec{H}}_{k,\mathrm{TX}}(t,\tau)\in\mathbb{C}^{J\times M}$ and $\tilde{\vec{H}}_{\mathrm{RX},k}(t,\tau)\in\mathbb{C}^{\tilde{M}\times J}$  in expression \eqref{eq:MIMO_TD_channel_general_SIM} can be expressed similarly to~\eqref{eq:MIMO_TD_channel_general_SIM_LOS} as $\tilde{P}$- and $\bar{P}$-path \ac{DD} \ac{MIMO} channels, respectively, namely:
\begin{equation}
\label{eq:MIMO_TD_channel_general_SIM_nLOS_TX}
\tilde{\vec{H}}_{k,\mathrm{TX}}(t,\tau) \triangleq \sqrt{\tfrac{JM}{\tilde{P}}} \sum_{\tilde{p}=1}^{\tilde{P}} h_{\tilde{p},k} e^{\jmath2\pi \nu_{\tilde{p},k} t} \delta( \tau - \tau_{\tilde{p},k}) \vec{B}_{\tilde{p},k},
%
%
\end{equation}
and
\begin{equation}
\label{eq:MIMO_TD_channel_general_SIM_nLOS_RX}
\tilde{\vec{H}}_{\mathrm{RX},k}(t,\tau) \triangleq \sqrt{\tfrac{J\tilde{M}}{\bar{P}}}  \sum_{\bar{p}=1}^{\bar{P}} h_{\bar{p},k} e^{\jmath 2\pi \nu_{\bar{p},k} t} \delta( \tau - \tau_{\bar{p},k}) \vec{B}_{\bar{p},k}.
%
%
\end{equation}

The notation 
$h_{\tilde{p},k}$ as well as $(\phi_{\tilde{p},k}^{\rm in},\theta_{\tilde{p},k}^{\rm in})$ and $(\phi_{\tilde{p},k}^{\rm out},\theta_{\tilde{p},k}^{\rm out})$ for the complex channel gain and pairs of azimuth and elevation \acp{AoA} and \acp{AoD}, respectively, for each $\tilde{p}$-th signal propagation path between the \ac{TX}-\ac{SIM} and each $k$-th \ac{RIS}, with $\tilde{p}=\{1,\ldots,\tilde{P}\}$ will be used henceforth. 
Similarly, $h_{\bar{p},k}$ as well as $(\phi_{\bar{p},k}^{\rm in},\theta_{\bar{p},k}^{\rm in})$ and $(\phi_{\bar{p},k}^{\rm out},\theta_{\bar{p},k}^{\rm out})$ indicate the complex channel gain and pairs of azimuth and elevation \acp{AoA} and \acp{AoD}, respectively, for each $\bar{p}$-th path between each $k$-th \ac{RIS} and the \ac{RX}-\ac{SIM}, with $\bar{p}=\{1,\ldots,\bar{P}\}$.
The delays and  Doppler shifts of the latter paths are denoted by $\tau_{\tilde{p},k}$, $\nu_{\tilde{p},k}$, and $\tau_{\bar{p},k}$, $\nu_{\bar{p},k}$, respectively, having similar bounds to $\tau_p$ and $\nu_p$ defined in Section~\ref{subsec:SISO_DD_Channel_Model}.

\begin{remark}[Special Cases]
Removing the second term with the $K$ summations in expression~\eqref{eq:MIMO_TD_channel_general_SIM} simplifies the full model to the \ac{MIMO} channel case including a \ac{TX} and \ac{RX} \ac{SIM} with no \acp{RIS} present.
In addition, removing any of the factors $\vec{\Upsilon}_\mathrm{R}(\tilde{\vec{\mathcal{Z}}})\vec{R}_\mathrm{RX}^{1/2}$ or $\vec{R}_\mathrm{TX}^{1/2}\vec{\Upsilon}_\mathrm{T}(\vec{\mathcal{Z}})$ in expression~\eqref{eq:prop_SIM_channel}, implies a \ac{MIMO} system lacking a \ac{RX}-\ac{SIM} or a \ac{TX}-\ac{SIM}, respectively. 
In the second of the latter two cases, $\tilde{\vec{H}}(\vec{\mathcal{F}},t,\tau)$ appearing in expressions~\eqref{eq:prop_SIM_channel} and \eqref{eq:MIMO_TD_channel_general_SIM} models signal propagation directly from the elements of the \ac{TX} \ac{ULA} (see the \ac{ULA} response vector in~\eqref{eq:ULA_response_vector_transmit}) to the \ac{RX}-\ac{SIM}, via the $K$ \acp{RIS}.
In this case, the size of $\tilde{\vec{H}}(\vec{\mathcal{F}},t,\tau)$ becomes $\tilde{M}\times N_\mathrm{T}$. 
Alternatively, when only the \ac{RX}-\ac{SIM} is missing, $\tilde{\vec{H}}(\vec{\mathcal{F}},t,\tau)$ will be of size $ N_\mathrm{R}\times M$. 
Finally, when none of the \ac{SIM} are considered, $\vec{H}(\vec{\mathcal{Z}},\tilde{\vec{\mathcal{Z}}},\vec{\mathcal{F}},t,\tau)\equiv\tilde{\vec{H}}(\vec{\mathcal{F}},t,\tau)$ representing the \acp{RIS}-empowered $N_\mathrm{R}\times N_\mathrm{T}$ \ac{DD} \ac{MIMO} channel. 
In the absence of \acp{RIS} and for a single-antenna \ac{TX} and \ac{RX} (i.e., $N_\mathrm{T}=N_\mathrm{R}=1$), the latter channel model reduces to the model described in \cite{Rou_SPM_2024} and summarized in Section~\ref{subsec:SISO_DD_Channel_Model}. 
\end{remark}

\section{Signal Modeling with OFDM, OTFS, and AFDM}
\label{sec:IO_Relationships}

In this section, the applicability of the proposed \ac{MPDD} \ac{MIMO} channel model is demonstrated by deriving various expressions for the \ac{TD} received signal corresponding to waveforms such as \ac{OFDM}, \ac{OTFS}, and \ac{AFDM}.

\subsection{Arbitrarily Modulated Signals}

Consider the deployment of fully digital beamformers in the \ac{MIMO} system showcased in Fig.~\ref{fig:system_model} at its $N_\mathrm{T}$-element \ac{TX} and $N_\mathrm{R}$-element \ac{RX}. Let $\vec{V} \in \mathbb{C}^{N_\mathrm{T} \times d_s}$ and $\vec{U} \in \mathbb{C}^{N_\mathrm{R} \times d_s}$ represent the transmit and receive digital beamformers, respectively, where $d_s \triangleq \text{min}(N_\mathrm{T},N_\mathrm{R})$ corresponds to the number of independent data streams to be communicated per coherent channel block. 
In addition, let $\vec{s}(t)$ represent the complex-valued and $d_s$-element transmit signal of any modulation (e.g., \ac{OFDM}, \ac{OTFS}, and \ac{AFDM}) in the \ac{TD}. Consequently, the $d_s$-element baseband received signal at a time instant $t$ (after the digital combiner) through the \ac{MPDD} \ac{MIMO} channel can be mathematically modeled as follows:
\begin{align}
    \label{eq:TD_I/O_relationship}
&\vec{r}(t) 
\triangleq  \vec{U}\herm \vec{H}(\vec{\vec{Z}},\tilde{\vec{\mathcal{Z}}},\vec{\mathcal{F}},t,\tau)  * \vec{V} \vec{s}(t) + \vec{w}(t)
\\&\!=\!\!\!\! \int\limits_{-\infty}^\infty \!\!\! \vec{U}\herm \vec{\Upsilon}_\mathrm{R}(\tilde{\vec{\mathcal{Z}}}) \vec{R}_\mathrm{RX}^{1/2} \bigg[ \tilde{\vec{H}}_d(t,\!\tau)\! +\!\!\! \sum_{k=1}^{K} \tilde{\vec{H}}_{\mathrm{RX},k}(t,\!\tau) \vec{\Phi}_k \tilde{\vec{H}}_{k,\mathrm{TX}}(t,\!\tau) \!\bigg] \vec{R}_\mathrm{TX}^{1/2} \vec{\Upsilon}_\mathrm{T}(\vec{\mathcal{Z}}) \vec{V} \vec{s}(t - \tau) d\tau, \nonumber
\end{align}
where $\vec{w}(t) \triangleq \vec{U}\herm \vec{n}(t) \in \mathbb{C}^{d_s \times 1}$ (omitted in the second inequality to avoid repetition) and $\vec{n}(t) \in \mathbb{C}^{N_\mathrm{R} \times 1}$ denotes the \ac{AWGN} vector at the RX side with spatially and temporally uncorrelated elements, each with zero mean and variance $\sigma_n^2$. 

Let $\vec{r}[n] \in \mathbb{C}^{d_s \times 1}$ and $\vec{s}[n] \in \mathbb{C}^{d_s \times 1}$, with $n \in \{ 0,\dots,N-1 \}$, be the finite sequences obtained after respectively sampling $\vec{r}(t)$ and $\vec{s}(t)$ at a sufficiently high sampling rate \cite{RanasingheICNCC2024} $F_S \triangleq \frac{1}{T_S}$ in Hz within a total bandwidth $B$. Then, the discrete-time equivalent of the received signal in expression~\eqref{eq:TD_I/O_relationship} can be obtained as follows:
\begin{align}
    \vec{r}[n]\! =\! \sum_{\ell=0}^\infty \bigg[ \bigg(&\sum_{k=1}^K  \sum_{\bar{p}=1}^{\bar{P}} \sum_{\tilde{p}=1}^{\tilde{P}} \overbrace{J \sqrt{\tfrac{\tilde{M} M}{\bar{P}\tilde{P}}} h_{\bar{p},k} h_{\tilde{p},k}  \vec{U}\herm \vec{\Upsilon}_\mathrm{R}(\tilde{\vec{\mathcal{Z}}})  \vec{R}_\mathrm{RX}^{1/2} \vec{B}_{\bar{p},k}
    \vec{\Phi}_k \vec{B}_{\tilde{p},k}
    \vec{R}_\mathrm{TX}^{1/2}  \vec{\Upsilon}_\mathrm{T}(\vec{\mathcal{Z}}) \vec{V}}^{\triangleq\check{\vec{H}}_{k,\bar{p},\tilde{p}}^\mathrm{RIS}(\vec{\mathcal{Z}},\tilde{\vec{\mathcal{Z}}},\vec{\Phi}_k,\vec{V},\vec{U}) \in \mathbb{C}^{d_s \times d_s}} \nonumber \\
    &\hspace{-11ex}\times e^{\jmath 2\pi\frac{n}{N} \overbrace{(f_{\bar{p},k}\! +\! f_{\tilde{p},k})}^{\triangleq\hat{f}_{k,\bar{p},\tilde{p}}}}  \delta[ \ell\! -\! \overbrace{(\ell_{\bar{p},k}\! +\! \ell_{\tilde{p},k})}^{\triangleq\hat{\ell}_{k,\bar{p},\tilde{p}}} ] \nonumber \\
    &\hspace{-11ex} + \sum_{p=1}^P \underbrace{\sqrt{\tfrac{M \tilde{M}}{P}} h_p \vec{U}\herm  \vec{\Upsilon}_\mathrm{R}(\tilde{\vec{\mathcal{Z}}})  \vec{R}_\mathrm{RX}^{1/2} \vec{B}_p
    \vec{R}_\mathrm{TX}^{1/2}  \vec{\Upsilon}_\mathrm{T}(\vec{\mathcal{Z}}) \vec{V}}_{\triangleq\check{\vec{H}}_p^d(\vec{\mathcal{Z}},\tilde{\vec{\mathcal{Z}}},\vec{V},\vec{U}) \in \mathbb{C}^{d_s \times d_s}}  e^{\jmath 2\pi f_p \frac{n}{N}}  \delta[ \ell\! -\! \ell_p ] \bigg)  \vec{s}[n\! -\! \ell] \bigg]\! + \!\vec{w}[n],
\label{eq:sampled_TD}
\end{align}
where $\ell$ indicates the normalized discrete delay index, while $f_p \triangleq \frac{N\nu_p}{F_s}$ and $\ell_p \triangleq \frac{\tau_p}{T_s}$ are respectively the normalized Doppler shift and the associated normalized discrete delay index of each $p$-th path propagation path between the \ac{TX}-\ac{SIM} and \ac{RX}-\ac{SIM}, with the definitions of $f_{\bar{p},k}$, $f_{\tilde{p},k}$, $\ell_{\bar{p},k}$, and $\ell_{\tilde{p},k}$ being similar for the respective channel paths. 

By taking into account a \ac{CP} of length $N_\mathrm{CP}$ and utilizing the circular convolution, the $N$-element discrete-time received signal in~\eqref{eq:sampled_TD} can be re-expressed as follows~\cite{Rou_SPM_2024}:
\begin{align}
    \vec{r}_v = \sum_{u=1}^{d_s} \!\bigg(\! \overbrace{\sum_{p=1}^P \check{h}_{p,v,u}^d  \underbrace{\vec{\Theta}_p  \vec{\Omega}^{f_p}  \vec{\Pi}^{\ell_p}}_{\triangleq\vec{G}_p \in \mathbb{C}^{N \times N}}}^{\triangleq\bar{\vec{H}}_{v,u}^d(\vec{\mathcal{Z}},\tilde{\vec{\mathcal{Z}}},\vec{V},\vec{U}) \in \mathbb{C}^{N \times N}} \!+\! \overbrace{\sum_{k=1}^{K}  \sum_{\bar{p}=1}^{\bar{P}} \sum_{\tilde{p}=1}^{\tilde{P}} \check{h}_{k,\bar{p},\tilde{p},v,u}^\mathrm{RIS} \overbrace{\vec{\Theta}_{k,\bar{p},\tilde{p}}  \vec{\Omega}^{\hat{f}_{k,\bar{p},\tilde{p}}}  \vec{\Pi}^{\hat{\ell}_{k,\bar{p},\tilde{p}}}}^{\triangleq\vec{G}_{k,\bar{p},\tilde{p}} \in \mathbb{C}^{N \times N}}}^{\triangleq\bar{\vec{H}}_{v,u}^\mathrm{RIS}(\vec{\mathcal{Z}},\tilde{\vec{\mathcal{Z}}},\vec{\mathcal{F}},\vec{V},\vec{U}) \in \mathbb{C}^{N \times N}}   \!\bigg)  \vec{s}_u \!+\! \vec{w}_v \nonumber \\
    &\hspace{-72.0ex}= \sum_{u=1}^{d_s} \!\underbrace{\Big(\bar{\vec{H}}_{v,u}^d(\vec{\mathcal{Z}},\tilde{\vec{\mathcal{Z}}},\vec{V},\vec{U}) \!+\! \bar{\vec{H}}_{v,u}^\mathrm{RIS}(\vec{\mathcal{Z}},\tilde{\vec{\mathcal{Z}}},\vec{\mathcal{F}},\vec{V},\vec{U})\Big)}_{\triangleq\bar{\vec{H}}_{v,u}^\mathrm{tot}(\vec{\mathcal{Z}},\tilde{\vec{\mathcal{Z}}},\vec{\mathcal{F}},\vec{V},\vec{U}) \in \mathbb{C}^{N \times N}}  \!\vec{s}_u \!+\! \vec{w}_v,
    \label{eq:vectorized_TD_IO}
\end{align}
where, for notational simplicity, the discrete-time index is omitted (also henceforth), which is implied.

In expression~\eqref{eq:vectorized_TD_IO}, the scalars $\check{h}_{p,v,u}^d$ and $\check{h}_{k,\bar{p},\tilde{p},v,u}^\mathrm{RIS}$, with $(v,u) = \{ 1,\dots,d_s \}$,  are the $(v,u)$-th elements of the matrices $\check{\vec{H}}_p^d(\vec{\mathcal{Z}},\tilde{\vec{\mathcal{Z}}},\vec{V},\vec{U})$ and $\check{\vec{H}}_{k,\bar{p},\tilde{p}}^\mathrm{RIS}(\vec{\mathcal{Z}},\tilde{\vec{\mathcal{Z}}},\vec{\Phi}_k,\vec{V},\vec{U})$, respectively, which are both implicitly defined in expression~\eqref{eq:sampled_TD}. In addition, $\vec{s}_u \triangleq [s_u[0],\ldots,s_u[N-1]] \in \mathbb{C}^{N \times 1}$ and $\vec{w}_v \triangleq [w_v[0],\ldots,w_v[N-1]] \in \mathbb{C}^{N \times 1}$ are the transmit signal and \ac{AWGN}  vectors for the $u$-th and $v$-th stream, respectively. In turn, each diagonal matrix $\vec{\Theta}_p \in \mathbb{C}^{N \times N}$ is defined as follows:
\begin{equation}
\label{eq:diagonal_CP_matrix_def}
    \vec{\Theta}_p \!\triangleq\! \text{diag}\Big( \![ \underbrace{e^{-\jmath 2\pi {\phi_\mathrm{CP}(\ell_p)}}\!, e^{-\jmath 2\pi {\phi_\mathrm{CP}(\ell_p - 1)}}\!, \dots, e^{-\jmath 2\pi {\phi_\mathrm{CP}(2)}}\!, e^{-\jmath 2\pi {\phi_\mathrm{CP}(1)}}}_{\ell_p \; \text{terms}}\!, \underbrace{1, \dots, 1}_{N - \ell_p \; \text{ones}}]\! \Big),
\end{equation}
capturing the effect of the \ac{CP} onto the $p$-th channel path, with $\phi_\mathrm{CP}(n)$ being a function of the sample index $n \in \{ 0,\ldots,N-1 \}$, representing a phase that depends on the specific waveform used. In addition, the diagonal matrix\footnote{Matrix exponentiation amounts to an element-wise exponentiation for diagonal matrices.} $\vec{\Omega} \in \mathbb{C}^{N \times N}$ given by
\begin{equation}
\label{eq:diagonal_Doppler_matrix_def}
    \vec{\Omega} \triangleq \text{diag}\Big([1,e^{-\jmath 2\pi /N},\dots,e^{-\jmath 2\pi (N-2) /N}, e^{-\jmath 2\pi (N-1) /N}]\Big) \in \mathbb{C}^{N \times N}
\end{equation}
contains the $N$ complex roots of unity, while $\vec{\Pi}\in \{0,1\}^{N \times N}$ is the forward cyclic shift matrix with elements defined as\footnote{The matrix $\vec{\Pi}$ is defined such that $\vec{A}\vec{\Pi}^{\ell_p}$, with $\ell_p\in\mathbb{N}$, is a cyclic left-shifted version of $\vec{A}$, $i.e.$, the first $\ell_p$ columns of $\vec{A}$ are moved to the positions of the last $\ell_p$ columns. It can be also seen that $\vec{\Pi}^0$ is the $N \times N$ identity matrix, yielding $\vec{A}\vec{\Pi}^0=\vec{A}$.}
\begin{equation}
\label{eq:PiMatrix}
    \pi_{i,j} \triangleq \delta_{i,j+1} + \delta_{i,j-(N-1)}\,\;\; \delta _{ij} \triangleq
    \begin{cases}
    0 & \text{if }i\neq j\\
    1 & \text{if }i=j
    \end{cases}.
    \vspace{-0.25ex}
\end{equation}

Leveraging the Kronecker product to concatenate all $d_s$ $\vec{r}_v$ vectors in expression~\eqref{eq:vectorized_TD_IO}, the following $N d_s$-element vector for the overall received signal in the \ac{TD}, considering an arbitrary modulated transmit signal, is obtained:
\begin{equation}
\vec{r}_\mathrm{TD} = \bar{\vec{H}}(\vec{\mathcal{Z}},\tilde{\vec{\mathcal{Z}}},\vec{\mathcal{F}},\vec{V},\vec{U})  \vec{s}_\mathrm{TD} + \bar{\vec{w}}_\mathrm{TD},
\label{eq:vectorized_TD_IO_kron}
\end{equation}
where $\bar{\vec{H}}(\vec{\mathcal{Z}},\tilde{\vec{\mathcal{Z}}},\vec{\mathcal{F}},\vec{V},\vec{U})\in \mathbb{C}^{N d_s \times N d_s}$ explicitly highlights the dependence of the \ac{TD} transfer function of the considered point-to-point \ac{MIMO} system on the \ac{TX} and \ac{RX} \ac{SIM}, the $K$ \acp{RIS} of the programmable smart wireless propagation environment, and the digital \ac{TX} and \ac{RX} beamformers, and is mathematically defined as follows:
\begin{eqnarray}
\bar{\vec{H}}(\vec{\mathcal{Z}},\tilde{\vec{\mathcal{Z}}},\vec{\mathcal{F}},\vec{V},\vec{U})\triangleq\sum_{p=1}^P (\check{\vec{H}}_p^d(\vec{\mathcal{Z}},\tilde{\vec{\mathcal{Z}}},\vec{V},\vec{U}) \otimes \vec{G}_p)&& \nonumber \\[-1ex]
&&\hspace{-26.9ex}+ \sum_{k=1}^{K}  \sum_{\bar{p}=1}^{\bar{P}} \sum_{\tilde{p}=1}^{\tilde{P}} (\check{\vec{H}}_{k,\bar{p},\tilde{p}}^\mathrm{RIS}(\vec{\mathcal{Z}},\tilde{\vec{\mathcal{Z}}},\vec{\Phi}_k,\vec{V},\vec{U}) \otimes \vec{G}_{k,\bar{p},\tilde{p}}),
\label{eq:H_bar}
\end{eqnarray}
with the $N d_s$-element vectors $\vec{s}_\mathrm{TD}$ and $\bar{\vec{w}}_\mathrm{TD}$ resulting from the concatenation of $\vec{s}_u$'s and $\vec{w}_v$'s in~\eqref{eq:vectorized_TD_IO}, respectively. For notational simplicity, the matrices $\check{\vec{H}}_p^d$ and $\check{\vec{H}}_{k,\bar{p},\tilde{p}}^\mathrm{RIS}$ appearing in equation \eqref{eq:H_bar} will hereafter be expressed without explicitly indicating their dependence on the \ac{TX}/\ac{RX} \ac{BF} and \ac{SIM}/\acp{RIS} parameters.

\subsection{OFDM Signaling}

Let $\mathcal{C}$ denote an arbitrary complex constellation set of cardinality $D$ and average energy $E_\mathrm{S}$, which is associated with a given digital modulation scheme (e.g., \ac{QAM}). In \ac{OFDM}, multiple information vectors $\vec{x}_u \in \mathcal{C}^{N\times 1}$ with $u = \{ 1,\dots,d_s \}$, containing a total of $Nd_s$ symbols, are modulated into the following transmit signal:
\begin{equation}
\label{eq:OFDM_modulation}
\vec{s}^{(\text{OFDM})}_u \triangleq \vec{F}_N\herm  \vec{x}_u \in \mathbb{C}^{N \times 1},
\end{equation}
where $\vec{F}_N$ denotes the $N$-point normalized \ac{DFT} matrix. 

After undergoing circular convolution with the \ac{DD} channel and using a formulation similar to expression~\eqref{eq:vectorized_TD_IO_kron}, the corresponding $Nd_s$-element discrete-time received \ac{OFDM} signal can be written as
\begin{equation}
\label{eq:TD_OFDM_input_output}
\vec{r}_\text{OFDM} \triangleq \bar{\vec{H}}(\vec{\mathcal{Z}},\tilde{\vec{\mathcal{Z}}},\vec{\mathcal{F}},\vec{V},\vec{U})  \vec{s}_\text{OFDM} + \bar{\vec{w}}_\mathrm{TD},
\end{equation}
where the $Nd_s$-element vectors are defined as
\begin{equation}
\label{eq:OFDM_stacked_s}
\vec{s}_\text{OFDM} \triangleq 
\begin{bmatrix}
\vec{s}^{(\text{OFDM})}_1 \\[-1ex]
\vdots \\
\vec{s}^{(\text{OFDM})}_{d_s}
\end{bmatrix},\,\,
\vec{r}_\text{OFDM} \triangleq 
\begin{bmatrix}
\vec{r}^{(\text{OFDM})}_1 \\[-1ex]
\vdots \\
\vec{r}^{(\text{OFDM})}_{d_s}
\end{bmatrix}.
\end{equation}

At the \ac{RX} side, applying \ac{OFDM} demodulation yields
\begin{equation}
\label{eq:OFDM_demodulation}
\vec{y}^{(\text{OFDM})}_v \triangleq \vec{F}_N  \vec{r}^{(\text{OFDM})}_v \in \mathbb{C}^{N \times 1},
\end{equation}
yielding the corresponding $Nd_s$-element discrete-time signal
\begin{equation}
\label{eq:OFDM_input_output}
\vec{y}_\text{OFDM} = \bar{\vec{H}}_\text{OFDM}(\vec{\mathcal{Z}},\tilde{\vec{\mathcal{Z}}},\vec{\mathcal{F}},\vec{V},\vec{U})  \vec{x} + \bar{\vec{w}}_\text{OFDM}, 
\end{equation}

\noindent where $\bar{\vec{w}}_\text{OFDM} \in \mathbb{C}^{Nd_s \times 1}$ is an equivalent \ac{AWGN} with the same statistics as $\bar{\vec{w}}_\mathrm{TD}$, and $\bar{\vec{H}}_\text{OFDM}(\vec{\mathcal{Z}},\tilde{\vec{\mathcal{Z}}},\vec{\mathcal{F}},\vec{V},\vec{U}) \in \mathbb{C}^{Nd_s \times Nd_s}$ represents the effective \ac{OFDM} channel defined similar to $\bar{\vec{H}}(\vec{\mathcal{Z}},\tilde{\vec{\mathcal{Z}}},\vec{\mathcal{F}},\vec{V},\vec{U})$ in~\eqref{eq:vectorized_TD_IO_kron}, which is expressed as
\begin{eqnarray}
\bar{\vec{H}}_\text{OFDM}
\triangleq \sum_{p=1}^P \check{\vec{H}}_p^d \otimes \overbrace{( \vec{F}_N \vec{G}_p  \vec{F}_N\herm)}^{\triangleq\vec{G}_p^\text{OFDM} \in \mathbb{C}^{N \times N}} + \sum_{k=1}^{K}  \sum_{\bar{p}=1}^{\bar{P}} \sum_{\tilde{p}=1}^{\tilde{P}} \check{\vec{H}}_{k,\bar{p},\tilde{p}}^\mathrm{RIS} \otimes \overbrace{( \vec{F}_N \vec{G}_{k,\bar{p},\tilde{p}}  \vec{F}_N\herm)}^{\triangleq\vec{G}_{k,\bar{p},\tilde{p}}^\text{OFDM} \in \mathbb{C}^{N \times N}}\nonumber \\
&&\hspace{-59ex}= \sum_{p=1}^P \check{\vec{H}}_p^d \otimes {\vec{G}_p^\text{OFDM}} + \sum_{k=1}^{K}  \sum_{\bar{p}=1}^{\bar{P}} \sum_{\tilde{p}=1}^{\tilde{P}} \check{\vec{H}}_{k,\bar{p},\tilde{p}}^\mathrm{RIS} \otimes {\vec{G}_{k,\bar{p},\tilde{p}}^\text{OFDM}}.
\label{eq:OFDM_effective_channel}
\end{eqnarray}
Notice that, for this \ac{OFDM} case, the \ac{CP} phase matrices $\vec{\Theta}_p$'s appearing in~\eqref{eq:vectorized_TD_IO} reduce to identity matrices \cite{Rou_SPM_2024}, i.e., $\phi_\mathrm{CP}(n) = 0$ in~\eqref{eq:diagonal_CP_matrix_def}, since there is no phase offset.

\subsection{OTFS Signaling}
When \ac{OTFS} is used, multiple matrices $\vec{X}_u\in \mathcal{C}^{\tilde{K}\times \tilde{K}'}$ with $u = \{ 1,\dots,d_s \}$, containing a total of $\tilde{K} \tilde{K}' d_s$ symbols taken from an arbitrary complex constellation $\mathcal{C}$, are modulated as follows\footnote{For simplicity, all pulse-shaping operations are assumed to utilize rectangular waveforms such that the corresponding sample matrices can be reduced to identity matrices.}:
\vspace{-0.5ex}
\begin{equation}
\label{eq:TD_transmit_matrix_vectorized}
\vec{s}^{(\text{OTFS})}_u \triangleq \text{vec}\big(\vec{S}_u\big) = (\vec{F}_{\tilde{K}'}\herm \otimes \vec{I}_{\tilde{K}})  \text{vec}\big( \vec{X}_u
\big) \in \mathbb{C}^{\tilde{K}\tilde{K}'\times 1},
\end{equation}
where $\text{vec}(\cdot)$ denotes matrix vectorization via column stacking and $\vec{S}_u$ is a \ac{TD} symbols' matrix obtained from\footnote{Equivalently, $\vec{S}_u$ can be obtained as the Heisenberg transform of the \ac{ISFFT} of $\vec{X}_u$, $i.e.$, $\vec{S}_u = \vec{F}_{\tilde{K}}\herm \vec{X}_\text{FT}^u$ with $\vec{X}_\text{FT}^u \triangleq \vec{F}_{\tilde{K}} \vec{X}_u \vec{F}_{\tilde{K}'}\herm \in \mathbb{C}^{\tilde{K}\times \tilde{K}'}$.} the \ac{IDZT} of $\vec{X}_u$ as follows~\cite{Hadani_WCNC_2017}:
\begin{equation}
\label{eq:TD_transmit_matrix}
\vec{S}_u = \vec{X}_u \vec{F}_{\tilde{K}'}\herm  \in \mathbb{C}^{\tilde{K}\times \tilde{K}'}.
\end{equation}
The notation in~\eqref{eq:TD_transmit_matrix_vectorized} is in line with the strategy described in \cite{Raviteja_TWC_2018}, whereby the \ac{OTFS} signals are first vectorized and then appended with a \ac{CP} of length $N_\mathrm{CP}$ in order to eliminate inter-frame interference, in similarity with \ac{OFDM}. 
Taking advantage of this similarity, and in order to allow for direct comparisons between the two waveforms, $\tilde{K}\times \tilde{K}' = N$ is set hereafter.

After transmission over the \ac{DD} channel $\bar{\vec{H}}(\vec{\mathcal{Z}},\tilde{\vec{\mathcal{Z}}},\vec{\mathcal{F}},\vec{V},\vec{U})$ as shown in expression~\eqref{eq:vectorized_TD_IO_kron}, the $Nd_s$-element discrete-time received \ac{OTFS} signal can be modeled similar to~\eqref{eq:TD_OFDM_input_output} as $\vec{r}_\text{OTFS} \triangleq \bar{\vec{H}}(\vec{\mathcal{Z}},\tilde{\vec{\mathcal{Z}}},\vec{\mathcal{F}},\vec{V},\vec{U})  \vec{s}_\text{OTFS} + \bar{\vec{w}}_\mathrm{TD}$, 
%
%
where the $Nd_s$-element vectors $\vec{s}_\text{OTFS}$ and $\vec{r}_\text{OTFS}$ are defined for \ac{OTFS} similar to expression~\eqref{eq:OFDM_stacked_s}.
However, unlike \ac{OFDM}, the detection of the information symbols $\vec{X}_u$'s from the $\vec{r}^{(\text{OTFS})}_v$ elements $\forall$$v=1,\ldots,d_s$ of $\vec{r}_\text{OTFS}$ requires reversing the vectorization and the \ac{IDZT} operations employed in the construction of the $d_s$ elements of $\vec{s}_\text{OTFS}$, resulting in a distinct effective channel. 
In particular, let $\vec{R}_v \triangleq \text{vec}^{-1}(\vec{r}^{(\text{OTFS})}_v) \in \mathbb{C}^{\tilde{K} \times \tilde{K}'}$, with $\text{vec}^{-1}(\cdot)$ indicating the de-vectorization operation according to which a vector of size $\tilde{K}\tilde{K}' \times 1$ is reshaped into a matrix of size $\tilde{K} \times \tilde{K}'$, and consider the following \ac{DZT}\footnote{Equivalently, $\vec{Y}_v$ can be obtained as the SFFT of the Wigner transform of $\vec{R}_v$: $\vec{Y}_\text{FT}^v \triangleq \vec{F}_{\tilde{K}} \vec{R}_v$, yielding $\vec{Y}_v = \vec{F}_{\tilde{K}}\herm \vec{Y}_\text{FT}^v \vec{F}_{\tilde{K}'}\in \mathbb{C}^{\tilde{K} \times \tilde{K}'}$.}
\begin{equation}
\label{eq:DD_rec_sig_after_SFFT}
\vec{Y}_v  =  \vec{R}_v \vec{F}_{\tilde{K}'} \in \mathbb{C}^{\tilde{K} \times \tilde{K}'}.
\end{equation}
The demodulated \ac{OTFS} signal at the \ac{RX} then becomes
\begin{equation}
\label{eq:DD_demodulation}
\vec{y}^{(\text{OTFS})}_v \triangleq \text{vec}(\vec{Y}_v) = (\vec{F}_{\tilde{K}'} \otimes \vec{I}_{\tilde{K}})  \vec{r}^{(\text{OTFS})}_v \in \mathbb{C}^{N\times 1},
\end{equation}
which can be compactly written, similar to~\eqref{eq:OFDM_input_output}, as the following $Nd_s$-element discrete-time received signal:
\begin{equation}
\label{eq:DD_input_output_relation}
\vec{y}_\text{OTFS} = \bar{\vec{H}}_\text{OTFS}(\vec{\mathcal{Z}},\tilde{\vec{\mathcal{Z}}},\vec{\mathcal{F}},\vec{V},\vec{U})  \vec{x} + \bar{\vec{w}}_\text{OTFS},
\end{equation}
where $\bar{\vec{w}}_\text{OTFS} \in \mathbb{C}^{Nd_s \times 1}$ is an equivalent \ac{AWGN} with the same statistics as $\bar{\vec{w}}_\mathrm{TD}$, while $\bar{\vec{H}}_\text{OTFS}(\vec{\mathcal{Z}},\tilde{\vec{\mathcal{Z}}},\vec{\mathcal{F}},\vec{V},\vec{U}) \in \mathbb{C}^{Nd_s \times Nd_s}$ represents the effective \ac{OTFS} channel and is given by

\begin{eqnarray}
\bar{\vec{H}}_\text{OTFS} \triangleq \sum_{p=1}^P \check{\vec{H}}_p^d \otimes \overbrace{( (\vec{F}_{\tilde{K}'} \otimes \vec{I}_{\tilde{K}}) \vec{G}_p  (\vec{F}_{\tilde{K}'}\herm \otimes \vec{I}_{\tilde{K}}))}^{\triangleq\vec{G}_p^\text{OTFS} \in \mathbb{C}^{N \times N}}&& \nonumber \\
&&\hspace{-35ex}+ \sum_{k=1}^{K}  \sum_{\bar{p}=1}^{\bar{P}} \sum_{\tilde{p}=1}^{\tilde{P}} \check{\vec{H}}_{k,\bar{p},\tilde{p}}^\mathrm{RIS} \otimes \overbrace{( (\vec{F}_{\tilde{K}'} \otimes \vec{I}_{\tilde{K}}) \vec{G}_{k,\bar{p},\tilde{p}} (\vec{F}_{\tilde{K}'}\herm \otimes \vec{I}_{\tilde{K}}))}^{\triangleq\vec{G}_{k,\bar{p},\tilde{p}}^\text{OTFS} \in \mathbb{C}^{N \times N}} \nonumber \\
&&\hspace{-37.3ex}= \sum_{p=1}^P \check{\vec{H}}_p^d \otimes {\vec{G}_p^\text{OTFS}} + \sum_{k=1}^{K}  \sum_{\bar{p}=1}^{\bar{P}} \sum_{\tilde{p}=1}^{\tilde{P}} \check{\vec{H}}_{k,\bar{p},\tilde{p}}^\mathrm{RIS} \otimes {\vec{G}_{k,\bar{p},\tilde{p}}^\text{OTFS}}.
\label{eq:OTFS_effective_channel}
\end{eqnarray}

Notice that similarly to the \ac{OFDM} case, the \ac{CP} phase matrices $\vec{\Theta}_p$'s reduce to identity matrices \cite{Rou_SPM_2024}.
Comparing the expressions in~\eqref{eq:OFDM_effective_channel} and~\eqref{eq:OTFS_effective_channel}, one can appreciate how \cite{Rou_SPM_2024}'s channel modeling approach elucidates both the similarity in form as well as the distinction in effect between the \ac{OFDM} and \ac{OTFS} waveforms in \ac{DD} channels, even under artificial enhancements such as the \acp{RIS} and \ac{SIM}.

\subsection{AFDM Signaling}

The signal for transmission per information vector $\vec{x}_u$ when \ac{AFDM} waveform is used for the considered \ac{DD} \ac{MIMO} channel is given by  the \ac{IDAFT}, as follows
\begin{equation}
\label{eq:AFDM_moduation}
\vec{s}^{(\text{AFDM})}_u \triangleq \vec{\Lambda}_1\herm  \vec{F}_{N}\herm  \vec{\Lambda}_2\herm  \vec{x}_u \in \mathbb{C}^{N \times 1},
\end{equation}
where the $N\times N$ matrices $\vec{\Lambda}_i$, with $i=1,2$, are defined as
\begin{equation}
\label{eq:lambda_def}
\vec{\Lambda}_i \triangleq \text{diag}\big(\big[1, e^{-\jmath2\pi c_i 2^2}, \ldots, e^{-\jmath2\pi c_i (N-1)^2}\big]\big),
\end{equation}
where the first central chirp frequency $c_1$ is an optimally designed parameter based on the maximum Doppler channel statistics \cite{Bemani_TWC_2023,Rou_SPM_2024}, while the second central chirp frequency $c_2$ is relatively a free parameter that can be exploited for \ac{ISAC} waveform shaping \cite{Zhu_Arxiv23} or information encoding \cite{Liu_Arxiv24,RouAsilimoar2024}.

It was shown in \cite{Rou_SPM_2024} that, after going through a \ac{DD} channel, an \ac{AFDM} modulated symbol vector $\vec{s}^{(\text{AFDM})}_u$ with the inclusion of a \ac{CPP} can be modeled similar to~\eqref{eq:vectorized_TD_IO}, by replacing the \ac{CP} matrix $\vec{\Theta}_p$ in expression~\eqref{eq:diagonal_CP_matrix_def} with the \ac{CPP} matrix generated via setting the function $\phi_\mathrm{CP}(n)$ in~\eqref{eq:diagonal_CP_matrix_def} to be $\phi_\mathrm{CP}(n) = c_1 (N^2 - 2Nn)$ \cite{KuranageTWC2024}. To this end, the $Nd_s$-element discrete-time received \ac{AFDM} signal can be modeled similar to~\eqref{eq:TD_OFDM_input_output} as $\vec{r}_\text{AFDM} \triangleq \bar{\vec{H}}(\vec{\mathcal{Z}},\tilde{\vec{\mathcal{Z}}},\vec{\mathcal{F}},\vec{V},\vec{U})  \vec{s}_\text{AFDM} + \bar{\vec{w}}_\mathrm{TD}$, where the $Nd_s$-element vectors $\vec{s}_\text{AFDM}$ and $\vec{r}_\text{AFDM}$ are defined for \ac{AFDM} as in~\eqref{eq:OFDM_stacked_s}. 

The \ac{AFDM} demodulation of each of the $\vec{r}^{(\text{AFDM})}_v$ with $v\in\{1,\ldots,d_s\}$ elements of $\vec{r}_\text{AFDM}$ is obtained as
\setcounter{equation}{38}
\begin{equation}
\vec{y}^{(\text{AFDM})}_v = \vec{\Lambda}_2  \vec{F}_{N}  \vec{\Lambda}_1  \vec{r}^{(\text{AFDM})}_v \in \mathbb{C}^{N\times 1},
\label{eq:AFDM_demodulation}
\end{equation}
yielding the following expression for the $Nd_s$-element discrete-time received signal (similar to expressions~\eqref{eq:OFDM_input_output} and~\eqref{eq:DD_input_output_relation}):
\begin{equation}
\vec{y}_\text{AFDM} = \bar{\vec{H}}_\text{AFDM}(\vec{\mathcal{Z}},\tilde{\vec{\mathcal{Z}}},\vec{\mathcal{F}},\vec{V},\vec{U})  \vec{x} + \bar{\vec{w}}_\mathrm{AFDM},
\label{eq:DAF_input_output_relation}
\end{equation}
where $\bar{\vec{w}}_\mathrm{AFDM} \in \mathbb{C}^{Nd_s \times 1}$ is an equivalent \ac{AWGN} holding the same statistics with $\bar{\vec{w}}_\mathrm{TD}$, and $\bar{\vec{H}}_\text{AFDM}(\vec{\mathcal{Z}},\tilde{\vec{\mathcal{Z}}},\vec{\mathcal{F}},\vec{V},\vec{U}) \in \mathbb{C}^{Nd_s \times Nd_s}$ indicates the effective \ac{AFDM} channel given by
\begin{eqnarray}
\bar{\vec{H}}_\text{AFDM} \triangleq \sum_{p=1}^P \check{\vec{H}}_p^d \otimes \overbrace{( \vec{\Lambda}_2  \vec{F}_{N}  \vec{\Lambda}_1 \vec{G}_p \vec{\Lambda}_1\herm  \vec{F}_{N}\herm  \vec{\Lambda}_2\herm)}^{\vec{G}_p^\text{AFDM} \in \mathbb{C}^{N \times N}}&& \nonumber \\
&&\hspace{-31.5ex}+ \sum_{k=1}^{K}  \sum_{\bar{p}=1}^{\bar{P}} \sum_{\tilde{p}=1}^{\tilde{P}} \check{\vec{H}}_{k,\bar{p},\tilde{p}}^\mathrm{RIS} \otimes \overbrace{( \vec{\Lambda}_2  \vec{F}_{N}  \vec{\Lambda}_1 \vec{G}_{k,\bar{p},\tilde{p}} \vec{\Lambda}_1\herm  \vec{F}_{N}\herm  \vec{\Lambda}_2\herm)}^{\vec{G}_{k,\bar{p},\tilde{p}}^\text{AFDM} \in \mathbb{C}^{N \times N}} \nonumber \\
&&\hspace{-33.5ex}=  \sum_{p=1}^P \check{\vec{H}}_p^d \otimes {\vec{G}_p^\text{AFDM}} + \sum_{k=1}^{K}  \sum_{\bar{p}=1}^{\bar{P}} \sum_{\tilde{p}=1}^{\tilde{P}} \check{\vec{H}}_{k,\bar{p},\tilde{p}}^\mathrm{RIS} \otimes {\vec{G}_{k,\bar{p},\tilde{p}}^\text{AFDM}}.
\label{eq:AFDM_effective_channel}
\end{eqnarray}
Clearly, the latter expression has the same structure of both~\eqref{eq:OFDM_effective_channel} and \eqref{eq:OTFS_effective_channel}, with the same holding for the \ac{MIMO} input-output relationships described by expressions~\eqref{eq:OFDM_input_output}, \eqref{eq:DD_input_output_relation}, and \eqref{eq:DAF_input_output_relation}. 
This implies that signal processing techniques such as channel estimation can be designed to under a unified framework, applying to \ac{OFDM}, \ac{OTFS}, \ac{AFDM}, and similar waveforms.

Finally, for the sake of clarity, a ``conventional'' \ac{DD}-\ac{MIMO} model -- $i.e.$, a \ac{MIMO} extension of the model in \cite{Rou_SPM_2024} without the incorporation of \ac{RIS} in the ambient and of the \ac{TX} and \ac{RX} \ac{SIM} corresponding to the \ac{MIMO} delay-time model defined in~\eqref{eq:MIMO_DD_channel_model} in Section~\ref{subsec:SISO_DD_Channel_Model} -- can be trivially extracted from the above.
For instance, for the \ac{OFDM}, \ac{OTFS}, and \ac{AFDM} waveforms, expressions \eqref{eq:OFDM_effective_channel}, \eqref{eq:OTFS_effective_channel}, and \eqref{eq:AFDM_effective_channel}, would yield
\begin{equation}
\label{eq:H_DD_MIMO}
\bar{\vec{H}}_\text{MIMO}\triangleq \sqrt{\tfrac{N_\mathrm{T} N_\mathrm{R}}{P}}\sum_{p=1}^P \left(h_p     \vec{a}_\mathrm{R}\left(\phi_p^{\rm in}\right) \vec{a}_\mathrm{T}\herm\left(\phi_p^{\rm out}\right)\right) \!\otimes\! {\vec{G}_p^\text{MIMO}},
\end{equation}
where the previous subscripts \ac{OFDM}, \ac{OTFS}, and \ac{AFDM}, are respectively represented by the generic subscript \ac{MIMO}.

\subsection{Comparison of Effective Channels}
\label{subsec:comp_eff_channel}

\begin{figure*}[t!]
    \centering
    \captionsetup[subfloat]{labelfont=small,textfont=small}
    \begin{minipage}[b]{.3\textwidth}
    \includegraphics[width=\textwidth]{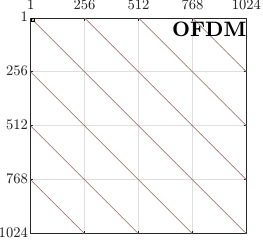}
    \end{minipage}
    \begin{minipage}[b]{.3\textwidth}
    \includegraphics[width=\textwidth]{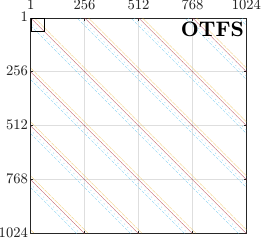}
    \end{minipage}
    \begin{minipage}[b]{.3\textwidth}
    \includegraphics[width=\textwidth]{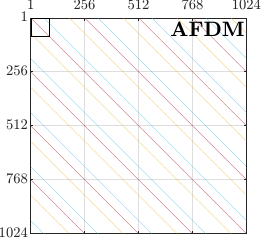}
    \end{minipage}
    \par\bigskip
    \vspace{-2ex}
    \subfloat[\label{fig:TX-SIM_to_RX-SIM_3D}
    Integer Doppler frequencies $(f_1, f_2, f_3) = (0, -2, 1)$.]{
    \begin{minipage}[b]{.3\textwidth}
    \includegraphics[width=\textwidth]{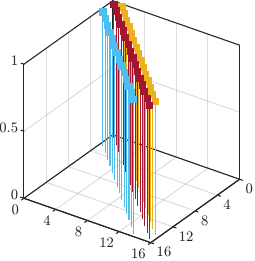}
    \end{minipage}
    \begin{minipage}[b]{.3\textwidth}
    \includegraphics[width=\textwidth]{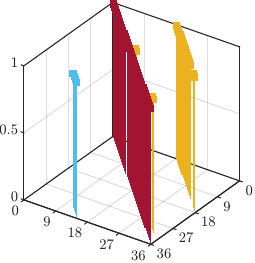}
    \end{minipage}
    \begin{minipage}[b]{.3\textwidth}
    \includegraphics[width=\textwidth]{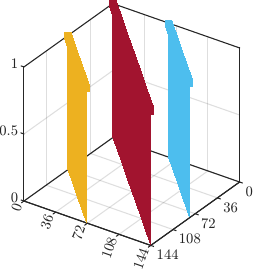}
    \end{minipage}  
    }%
    \par\bigskip
    \vspace{-2ex}
    \begin{minipage}[b]{.3\textwidth}
    \includegraphics[width=\textwidth]{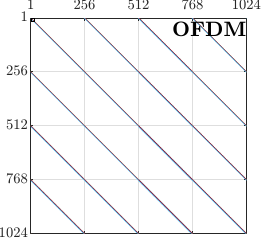}
    \end{minipage}
    \begin{minipage}[b]{.3\textwidth}
    \includegraphics[width=\textwidth]{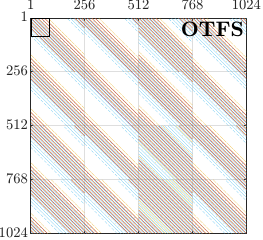}
    \end{minipage}
    \begin{minipage}[b]{.3\textwidth}
    \includegraphics[width=\textwidth]{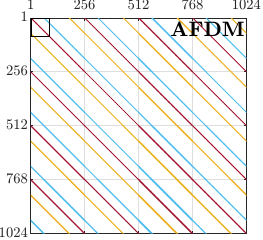}
    \end{minipage}  
    \par\bigskip
    \vspace{-1ex}
    \subfloat[\label{fig:TX-SIM_to_RX-SIM_frac_3D}
    Fractional Doppler frequencies $(f_1, f_2, f_3) = (0.698, -1.477, 1.124)$.]{
    \begin{minipage}[b]{.3\textwidth}
    \includegraphics[width=\textwidth]{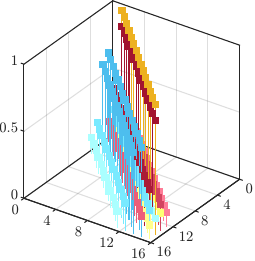}
    \end{minipage}
    \begin{minipage}[b]{.3\textwidth}
    \includegraphics[width=\textwidth]{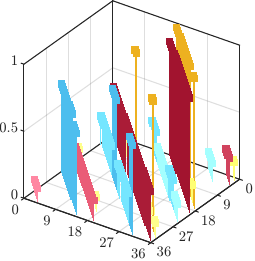}
    \end{minipage}
    \begin{minipage}[b]{.3\textwidth}
    \includegraphics[width=\textwidth]{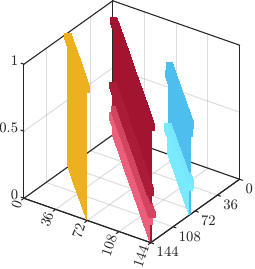}
    \end{minipage}  
    }%
    \caption{Unoptimized $4\times4$ \ac{MPDD}-\ac{MIMO} with identical \ac{TX} and \ac{RX} \ac{SIM} with  $Q = \tilde{Q} = 5$ layers and $M = \tilde{M} = 100$ meta-atoms per layer, considering \ac{OFDM}, \ac{OTFS}, and \ac{AFDM} with $N=256$ symbols per frame and $P = 3$ channel paths with respective delays $(\ell_1, \ell_2, \ell_3) = (0, 5, 14)$ and integer (figure a) as well as fractional (figure b) Doppler frequencies.
    The 3D versions show the amplitude of the channel taps corresponding to the carriers outlined by the squares of the upper left corners.}
\label{fig:TX-SIM_to_RX-SIM_Full}
\vspace{-3ex}
\end{figure*}

As an example of how the aforementioned waveforms are affected by a MPDD \ac{MIMO} channel, consider a point-to-point system with $N_\mathrm{T} = N_\mathrm{R} = 4$ antennas, such that up to $d_s = 4$ independent data streams can be utilized.
Assume that the system operates with a carrier frequency of $28$ GHz ($i.e.$, at wavelength $\lambda = 10.7$ mm) with a bandwidth $B = 20$ MHz in an environment with identical \ac{TX} and \ac{RX} \ac{SIM}, both with $Q = \tilde{Q} = 5$ layers of metasurfaces and $M = \tilde{M} = 100$ meta-atoms per layer, leading to $M_x = M_z = \tilde{M}_x = \tilde{M}_z = 10$. 
In line with the Rayleigh-Sommerfeld diffraction theory, the distance between any two adjacent metasurface layers at either the \ac{TX} or \ac{RX} is $5\lambda$, while the distance between two adjacent meta-atoms on any layer is $\lambda/2$ both along the $x$- and $z$-axis, such that $\rho_t = \rho_r = \lambda^2/4$. 
The assumptions that the \acp{SIM} are unoptimized and that the environment lacks any \ac{RIS} presence are also adopted, thus, we set $\vec{\mathcal{Z}} = \tilde{\vec{\mathcal{Z}}} = \vec{\mathcal{F}} = \vec{I}_{M/\tilde{M}/J}$ $\forall q, \tilde{q}, k$.

Since the \acp{ULA} are aligned with the meta-atoms on the \ac{SIM}, the angle between the propagation and the normal direction of the metasurface layers becomes in this case $\epsilon^{q}_{m,m'} = \tilde{\epsilon}^{\tilde{q}}_{\tilde{m},\tilde{m}'} = 0$ $\forall q,\tilde{q},m,\tilde{m},m',\tilde{m}'$. Consequently, $\vec{\Psi}_q$'s, $\vec{\Gamma}_q$'s, $\vec{\Delta}_q$'s, and $\vec{\Xi}_q$'s can be generated from expressions \eqref{eq:diagona_shift_matrix_per_layer} and \eqref{eq:diffraction_coeff}, leading to the generation of the transfer functions $\vec{\Upsilon}_\mathrm{T}$ and $\vec{\Upsilon}_\mathrm{R}$ for the \ac{TX} and \ac{RX} \ac{SIM}, respectively, as given in~\eqref{eq:transmit_SIM_full} and \eqref{eq:receive_SIM_full}, respectively. 

The \ac{TX}/\ac{RX} digital beamformers are also unoptimized and set as $\vec{V} = \vec{U} = \vec{I}_{d_s}$. Finally, the sampling frequency is set to $F_\mathrm{S}=B$ and the number of symbols per frame to $N = 256$. For the \ac{MPDD} channel, the path delays $\tau_p$'s are uniformly distributed in $[0,\tau_\text{max}]$, and that the Doppler shifts follow a Jakes spectrum, $i.e.$, $\nu_p = \nu_\text{max} \cos(\theta_p)$ $\forall p$ with each $\theta_p$ uniformly distributed in $[-\pi,\pi]$. 
In turn, the \ac{2D} and \ac{3D} elevation \acp{AoD}/\acp{AoA} are uniformly distributed in $[0,\pi]$, while the \ac{3D} azimuth \acp{AoD}/\acp{AoA} are in $[-\frac{\pi}{2},\frac{\pi}{2}]$. 
Finally, the setup considers a case with $P = 3$ paths, with respective delays $[\ell_1, \ell_2, \ell_3] = [0, 5, 14]$. 

Figure \ref{fig:TX-SIM_to_RX-SIM_Full} illustrates the resulting unoptimized \ac{MPDD}-\ac{MIMO} channels with all considered waveforms. 
As observed from Fig. \ref{fig:TX-SIM_to_RX-SIM_frac_3D}, the fractional components of the Doppler shift ``spread'' the path-wise components, making the effective channel matrix design vital for estimation/detection tasks to avoid overlaps and mixing among paths.
Comparing the results shown in Fig.~\ref{fig:TX-SIM_to_RX-SIM_Full} with those in \cite{Rou_SPM_2024}, it can be confirmed that, as expected, unoptimized \ac{SIM} have no effect onto the \ac{DD} channels undergone by the compared \ac{OFDM}, \ac{OTFS}, and \ac{AFDM} waveforms.
In what follows, however, it is demonstrated that when optimized, these structures can significantly impact on the detection performance of such systems.

\section{SIM Optimization and Data Detection}
\label{sec:SIM_Optimization_Detection}

A great potential advantage of \ac{MIMO} systems incorporating \ac{SIM}, compared to conventional \ac{MIMO} systems, is that signal processing functions previously carried out by circuitry, be it in analog or digital fashions, can instead be performed in the wave propagation domain~\cite{AnJSAC2023,wave_comp_2}.
And while this new wave-domain processing capability can be exploited to replace classic digital/analog processing, as suggested $e.g.$ in \cite{AnWC2024,wave_comp_1}, it can also be utilized to augment it.
Focusing on the latter case, and to elaborate further, consider, for example the trade-off that exists between enhancing the \ac{SNR} of received signals and improving communication rate \cite{TseTIT2004}, which from the viewpoint of receiver design 
translates to either designing directivity-enhancing receive beamformers ($i.e.$, combiners) or, instead, exploiting the multiple streams of data as extrinsic information for detection \cite{MolischCOMMMAG2017}. Under the conventional paradigm, a choice (or trade-off) between these competing interests must be made.
In contrast, in the case of \ac{MIMO}-\ac{SIM} systems, one can seek to reap the best of both worlds, by parameterizing the \ac{SIM} for \ac{SNR} gain, while leaving the degrees of freedom afforded by the multiple \ac{RX} antennas to design robust detectors.


With the latter approach in mind, this section offers an illustrative application of the channel model detailed above.
In particular, an optimization problem that leverages the reconfigurability of the proposed \ac{MPDD}-\ac{MIMO} model to increase the intensity of the complex channel coefficients at the \ac{EM} domain is formulated,
and subsequently, introduce a GaBP-based data detection algorithm that exploits the signals from all \ac{RX} antennas simultaneously and extrinsically~\cite{MPDD_MIMO}. The flexibility of the model is highlighted by offering both of the aforementioned contributions in a manner that they apply to \ac{OFDM}, \ac{OTFS} and \ac{AFDM} alike.

\subsection{SIM-Based Signal Enhancement}
\label{SIM_Optimization}

Referring to expression~\eqref{eq:sampled_TD}, setting the \ac{BF} matrices $\vec{U}$ and $\vec{V}$ to identities to put emphasis on the impact of \ac{SIM} as per the discussion above, and considering for simplicity a scenario without any \ac{RIS}, the goal is to parametrize the matrices $\vec{\mathcal{Z}}$ and $\tilde{\vec{\mathcal{Z}}}$ to enhance the total receive signal power, which can be achieved by solving the following optimization problem:
\begin{align}
\label{eq:full_optimization_problem_channel_coeff}
&\underset{\vec{\mathcal{Z}},\tilde{\vec{\mathcal{Z}}}}{\text{max}} \; \mathcal{O}(\vec{\mathcal{Z}},\tilde{\vec{\mathcal{Z}}}) = \sum_{p=1}^P \Big|\Big| \overbrace{\tilde{h}_p \vec{\Upsilon}_\mathrm{R}(\tilde{\vec{\mathcal{Z}}}) \vec{R}_\mathrm{RX}^{1/2} \vec{B}_p \vec{R}_\mathrm{TX}^{1/2} \vec{\Upsilon}_\mathrm{T}(\vec{\mathcal{Z}})}^{\triangleq\,\vec{O}_p}  \Big|\Big|_F^2\nonumber\\
&\;\;\;\text{\text{s}.\text{t}.}\;\vec{\Upsilon}_\mathrm{T}(\vec{\mathcal{Z}})\;\text{as in eq. \eqref{eq:transmit_SIM_full}},\;\vec{\Upsilon}_\mathrm{R}(\tilde{\vec{\mathcal{Z}}})\;\text{as in eq. \eqref{eq:receive_SIM_full}},\nonumber\\
&\;\;\;\quad\;\;\,\vec{\Psi}_q \; \text{and} \; \vec{\Delta}_{\tilde{q}}\;\text{as in eqs. \eqref{eq:diagona_shift_matrix_per_layer} and \eqref{eq:diagona_shift_matrix_per_layer_rx}, respectively},\nonumber\\
&\;\;\;\quad\;\;\,|\zeta^{q}_{m}| \leq \pi \; \forall (q,m), \text{and\;} \,|\tilde{\zeta}^{\tilde{q}}_{\tilde{m}}| \leq \pi \; \forall (\tilde{q},\tilde{m}).
\end{align}
Notice that the array response matrix $\vec{B}_p$ corresponding to a $p$-th path, defined previously in~\eqref{eq:Bmatrix}, as well as the scalar $\tilde{h}_p$ defined as $\tilde{h}_p \triangleq h_p \sqrt{\tfrac{M \tilde{M}}{P}}$ were used here.
%

In view of the non-convex unit modulus constraints in~\eqref{eq:full_optimization_problem_channel_coeff}'s optimization problem, a simple gradient ascent technique is utilized to tune the phase shift parameters of the \ac{SIM}.
In particular, following \cite{AnICC2024}, the gradient ascent algorithm is employed to adjust the phase shifts of the transmit and receive \ac{SIM} iteratively, maximizing the objective function in equation \eqref{eq:full_optimization_problem_channel_coeff}. 
The algorithm is divided into two main discrete steps, addressed in detailed below.

\subsubsection{Gradient Calculation} The gradient of the objective function $\mathcal{O}(\vec{\mathcal{Z}},\tilde{\vec{\mathcal{Z}}})$ with respect to the phase shift vector of the $q$-th layer of the \ac{TX}-\ac{SIM}, denoted by $\vec{\zeta}_{q} = [\zeta^{q}_{1}, \zeta^{q}_{2}, \dots, \zeta^{q}_{M}]\trans $, is given by the following expression $\forall q$:
\begin{equation}
\label{eq:gradient_def_tx}
\nabla_{\vec{\zeta}_{q}} \mathcal{O}(\vec{\mathcal{Z}},\tilde{\vec{\mathcal{Z}}}) = \sum_{p=1}^P \sum_{n_t=1}^{N_T} \nabla_{\vec{\zeta}_{q}} ||\vec{o}_{p,n_t}||^2,
\end{equation}
where $\vec{o}_{p,n_t} \in \mathbb{C}^{N_R \times 1}$, with $n_t = \{1, 2, \dots, N_T\}$ and $p  = \{1, 2, \dots, P\}$, represents the $n_t$-th column of the $p$-th $\vec{O}_{p}$ matrix defined within the objective function in problem~\eqref{eq:full_optimization_problem_channel_coeff}. Similarly, the gradient with respect to the phase shift vector $\tilde{\vec{\zeta}}_{\tilde{q}} = [ \tilde{\zeta}^{\tilde{q}}_{1}, \tilde{\zeta}^{\tilde{q}}_{2}, \dots, \tilde{\zeta}^{\tilde{q}}_{\tilde{M}} ]\trans$ of the $\tilde{q}$-th layer of the \ac{RX}-\ac{SIM} is given by
\vspace{-1ex}
\begin{equation}
\label{eq:gradient_def_rx}
\nabla_{\tilde{\vec{\zeta}}_{\tilde{q}}} \mathcal{O}(\vec{\mathcal{Z}},\tilde{\vec{\mathcal{Z}}}) = \sum_{p=1}^P \sum_{n_t=1}^{N_T} \nabla_{\tilde{\vec{\zeta}}_{\tilde{q}}} ||\vec{o}_{p,n_t}||^2, \; \forall \tilde{q}.
\end{equation}

Leveraging the chain rule, the per-shift partial derivates of $||\vec{o}_{p,n_t}||^2$ with respect to $\zeta^{q}_{m}$ are given as follows $\forall m,q$:
\begin{align}
\label{eq:partial_tx_o}
\frac{\partial ||\vec{o}_{p,n_t}||^2}{\partial \zeta^{q}_{m}} & = 2 \Re \Bigg\{ \frac{\partial \vec{o}_{p,n_t} \herm }{ \partial \zeta^{q}_{m} }  \vec{o}_{p,n_t} \Bigg\} = 2 \Re \Bigg\{ \frac{\partial \big( \tilde{\vec{\Upsilon}}_{t:q,p,n_t}  \vect{\vec{\Psi}_q} \big) \herm }{ \partial \zeta^{q}_{m} }  \vec{o}_{p,n_t} \Bigg\} \nonumber \\
& = 2 \Re \Big\{ -\jmath e^{-\jmath \zeta^{q}_{m}} \vec{i}_{m}\trans \tilde{\vec{\Upsilon}}_{t:q,p,n_t}\herm  \vec{o}_{p,n_t} \Big\} = 2 \Im \Big\{ e^{-\jmath \zeta^{q}_{m}} \vec{i}_{m}\trans \tilde{\vec{\Upsilon}}_{t:q,p,n_t}\herm  \vec{o}_{p,n_t} \Big\},
\end{align}
where $\vec{i}_m$ stands for the $m$-th column of $\vec{I}_{M}$ and the second equality holds due to the fact that $\vec{o}_{p,n_t} = \tilde{\vec{\Upsilon}}_{t:q,p,n_t} \vect{\vec{\Psi}_{q}}$, with $\tilde{\vec{\Upsilon}}_{t:q,p,n_t} \in \mathbb{C}^{N_R \times M}$ denoting the equivalent channel matrix of the $p$-th path associated to the $q$-th layer of the \ac{TX}-\ac{SIM} and the $n_t$-th transmit antenna, which is defined as follows
\begin{align}
\tilde{\vec{\Upsilon}}_{t:q,p,n_t} \triangleq& \tilde{h}_p \vec{\Upsilon}_\mathrm{R}(\tilde{\vec{\mathcal{Z}}}) \vec{R}_\mathrm{RX}^{1/2} \vec{B}_p \vec{R}_\mathrm{TX}^{1/2} \prod_{q'=1}^{q+1} \vec{\Psi}_{Q-q'+1} \vec{\Gamma}_{Q-q'+1} \text{diag}(\vec{s}_{q,n_t}),
\end{align}
where $\vec{s}_{q,n_t} \in \mathbb{C}^{M \times 1}$ is the signal component activating the $q$-th layer of the \ac{TX}-\ac{SIM} associated to the $n_t$-th transmit antenna, which is defined as the $n_t$-th column of 
\begin{align}
\vec{S}_{q} = \vec{\Gamma}_{q}\!\!\!\!\!\!\! \prod_{q'= Q - q + 2}^{Q}\!\!\!\!\!\!\! \vec{\Psi}_{Q-q'+1} \vec{\Gamma}_{Q-q'+1}.
\end{align}
Finally, the $M$ partial derivatives of each $p$-th path can be gathered into a vector, yielding the following final expression for the gradient of the \ac{TX}-\ac{SIM}:
\begin{equation}
\label{eq:final_grad_TXSIM}
\nabla_{\vec{\zeta}_{q}} \mathcal{O}(\vec{\mathcal{Z}},\tilde{\vec{\mathcal{Z}}}) = 2 \Im \Bigg\{ \sum_{p=1}^P \sum_{n_t=1}^{N_T} \vec{\Psi}_{q}\herm \tilde{\vec{\Upsilon}}_{t:q,p,n_t}\herm \vec{o}_{p,n_t} \Bigg\}.
\end{equation}

Similarly, considering the definition of the gradient for the \ac{RX}-\ac{SIM} in~\eqref{eq:gradient_def_rx} and expressing the per-shift partial derivates with respect to $\zeta^{\tilde{q}}_{\tilde{m}}$, yields $\forall \tilde{m},\tilde{q}$:
\begin{align}
\label{eq:partial_rx_o}
\frac{\partial ||\vec{o}_{p,n_t}||^2}{\partial \zeta^{\tilde{q}}_{\tilde{m}}} & = 2 \Re \Bigg\{ \frac{\partial \vec{o}_{p,n_t} \herm }{ \partial \zeta^{\tilde{q}}_{\tilde{m}} }  \vec{o}_{p,n_t} \Bigg\} = 2 \Re \Bigg\{ \frac{\partial \big( \tilde{\vec{\Upsilon}}_{r:\tilde{q},p,n_t}  \vect{\tilde{\vec{\Delta}}_{\tilde{q}}} \big) \herm }{ \partial \zeta^{\tilde{q}}_{\tilde{m}} }  \vec{o}_{p,n_t} \Bigg\} \nonumber \\
& = 2 \Re \Big\{ -\jmath e^{-\jmath \zeta^{\tilde{q}}_{\tilde{m}}} \vec{i}_{\tilde{m}}\trans \tilde{\vec{\Upsilon}}_{r:\tilde{q},p,n_t}\herm  \vec{o}_{p,n_t} \Big\} = 2 \Im \Big\{ e^{-\jmath \zeta^{\tilde{q}}_{\tilde{m}}} \vec{i}_{\tilde{m}}\trans \tilde{\vec{\Upsilon}}_{r:\tilde{q},p,n_t}\herm  \vec{o}_{p,n_t} \Big\}, 
\end{align}
where $\vec{i}_{\tilde{m}}$ stands for the $\tilde{m}$-th column of $\vec{I}_{\tilde{M}}$ and the second equality holds due to $\vec{o}_{p,n_t} = \tilde{\vec{\Upsilon}}_{r:\tilde{q},p,n_t} \vect{\vec{\Delta}_{\tilde{q}}}$, with $\tilde{\vec{\Upsilon}}_{r:\tilde{q},p,n_t} \in \mathbb{C}^{N_R \times \tilde{M}}$ denoting the equivalent channel matrix of the $p$-th path associated to the $\tilde{q}$-th layer of the RX-\ac{SIM} and the $n_t$-th transmit antenna, which is defined as follows:
\begin{equation}
\tilde{\vec{\Upsilon}}_{r:\tilde{q},p,n_t} \triangleq \tilde{h}_p \vec{\Xi}_{1} \Bigg( \prod_{\tilde{q}'=1}^{\tilde{q} - 1} \vec{\Delta}_{\tilde{q}'} \vec{\Xi}_{\tilde{q}' + 1} \Bigg) \text{diag}(\tilde{\vec{s}}_{\tilde{q},n_t}).
\end{equation}
In this expression, $\tilde{\vec{s}}_{\tilde{q},n_t} \in \mathbb{C}^{\tilde{M} \times 1}$ is the signal component activating the $\tilde{q}$-th layer of the \ac{RX}-\ac{SIM} associated to the $n_t$-th transmit antenna, which is defined as the $n_t$-th column of the following matrix:
\begin{equation}
\tilde{\vec{S}}_{\tilde{q}} = \Bigg( \prod_{\tilde{q}'= \tilde{q} + 1}^{\tilde{Q}} \vec{\Xi}_{\tilde{q}'} \vec{\Delta}_{\tilde{q}'} \Bigg) \vec{R}_\mathrm{RX}^{1/2} \vec{B}_p \vec{R}_\mathrm{TX}^{1/2} \vec{\Upsilon}_\mathrm{T}(\vec{\mathcal{Z}}).
\end{equation}
Finally, the $\tilde{M}$ partial derivatives of each $p$-th path can be gathered, giving the final calculation for the gradient of the \ac{TX}-\ac{SIM} as follows:
\begin{equation}
\label{eq:final_grad_RXSIM}
\nabla_{\tilde{\vec{\zeta}}_{\tilde{q}}} \mathcal{O}(\vec{\mathcal{Z}},\tilde{\vec{\mathcal{Z}}}) = 2 \Im \Bigg\{ \sum_{p=1}^P \sum_{n_t=1}^{N_T} \vec{\Delta}_{\tilde{q}}\herm \tilde{\vec{\Upsilon}}_{r:\tilde{q},p,n_t}\herm \vec{o}_{p,n_t} \Bigg\}.
\end{equation}

\subsubsection{Parameter Update} 
With the aforementioned closed-form expressions for the gradient of $\mathcal{O}(\vec{\mathcal{Z}},\tilde{\vec{\mathcal{Z}}})$ with respect to the \ac{TX}- and \ac{RX}-\ac{SIM} phase parameters given respectively by the expressions~\eqref{eq:final_grad_TXSIM} and \eqref{eq:final_grad_RXSIM},
the update required to iteratively adjust the phases $\vec{\zeta}_{q}$ and $\tilde{\vec{\zeta}}_{\tilde{q}}$ to optimize the total receive powers, as described in the optimization problem~\eqref{eq:full_optimization_problem_channel_coeff}, can be efficiently computed by 
\begin{subequations}
\label{eq:param_updates_GD}
\begin{equation}
    \vec{\zeta}_{q}^{(i + 1)} = \vec{\zeta}_{q}^{(i)} + \lambda^{i} \rho^{(i)} \nabla_{\vec{\zeta}_{q}} \mathcal{O}(\vec{\mathcal{Z}},\tilde{\vec{\mathcal{Z}}}),
\end{equation}
\begin{equation}
    \tilde{\vec{\zeta}}_{\tilde{q}}^{(i + 1)} = \tilde{\vec{\zeta}}_{\tilde{q}}^{(i)} + \lambda^{i} \tilde{\rho}^{(i)} \nabla_{\tilde{\vec{\zeta}}_{\tilde{q}}} \mathcal{O}(\vec{\mathcal{Z}},\tilde{\vec{\mathcal{Z}}}),
\end{equation}
\end{subequations}
where $\lambda \in (0,1)$ is the decaying learning rate parameter to ensure convergence and $\rho^{(i)},\tilde{\rho}^{(i)}$ are normalization parameters calculated at each step as follows:
\begin{subequations}
\label{eq:normalization_params}
\begin{equation}
    \rho^{(i)} = \pi / \underset{q \in Q, m \in M}{\max}\nabla_{\vec{\zeta}_{q}} \mathcal{O}(\vec{\mathcal{Z}},\tilde{\vec{\mathcal{Z}}}),
\end{equation}
\begin{equation}
    \tilde{\rho}^{(i)} = \pi / \underset{\tilde{q} \in \tilde{Q}, \tilde{m} \in \tilde{M}}{\max}\nabla_{\tilde{\vec{\zeta}}_{\tilde{q}}} \mathcal{O}(\vec{\mathcal{Z}},\tilde{\vec{\mathcal{Z}}}).
\end{equation}
\end{subequations}

\subsection{GaBP-Based Data Detection}
In possession of the \ac{SIM} optimization method detailed above, and given the  input-output relationships given in Section~\ref{sec:IO_Relationships} for various exemplary \ac{ISAC}-enabling waveforms, the impact of integrating \ac{SIM} onto the design of communication systems under the \ac{MPDD}-\ac{MIMO} channel model described in Section~\ref{sec:MPDD_MIMO_Channel_Model} can be demonstrated, by comparing the corresponding performances of \ac{OFDM}, \ac{OTFS} and \ac{AFDM}, with and without \ac{SIM}.

Before proceeding, recall that it has been widely demonstrated \cite{GaudioTWC2020,KuranageTWC2024} that \ac{AFDM} and \ac{OTFS} can significantly outperform \ac{OFDM} under \ac{DD} conditions.
It shall be demonstrated, however, that \ac{SIM} can significantly lower the performance gap between these waveforms which in turn, given the potential of the technology to reduce hardware complexity compared to traditional digital signal processing techniques, suggests that the design of waveforms to combat \ac{DD} distortion can be significantly impacted by the emergence of \ac{SIM}.

From a receiver design viewpoint, the aim is to estimate the transmitted signal $\vec{x}$, under the assumption that the effective channel matrix $\bar{\vec{H}}$ is known. In order to derive a \ac{GaBP}-based detector for arbitrary waveforms, consider the following generic \ac{I/O} relationship:
\begin{equation}
\label{General_I/O_arbitrary}
\vec{y} = \bar{\vec{H}}  \vec{x} + \bar{\vec{w}}, 
\end{equation}
where, for conciseness, the waveform specific subscripts are dropped from the notation of the effective channel matrix $\bar{\vec{H}}$, which for \ac{OFDM}, \ac{OTFS}, and \ac{AFDM} are respectively given by expressions~\eqref{eq:OFDM_effective_channel}, \eqref{eq:OTFS_effective_channel}, and \eqref{eq:AFDM_effective_channel}. By setting $\bar{N} = \bar{M} \triangleq N d_s \times N d_s$ with $\bar{n} \triangleq \{1,\dots,\bar{N}\}$ and $\bar{m} \triangleq \{1,\dots,\bar{M}\}$, the element-wise relationship corresponding to~\eqref{General_I/O_arbitrary} is given by
\begin{equation}
\label{General_I/O_arbitrary_elementwise}
y_{\bar{n}} = \sum_{\bar{m}=1}^{\bar{M}} \bar{h}_{\bar{n},\bar{m}} x_{\bar{m}} + \bar{w}_{\bar{n}}, 
\end{equation}
such that the soft replica of the $\bar{m}$-th communication symbol associated with the $\bar{n}$-th receive signal $y_{\bar{n}}$, computed at the $i$-th iteration of a message-passing algorithm, can be denoted by $\hat{x}_{\bar{n},\bar{m}}^{(i)}$, with the corresponding \ac{MSE} of these estimates computed for the $i$-th iteration as follows $\forall \bar{n},\bar{m}$:
\begin{equation}
\hat{\sigma}^{2(i)}_{x:{\bar{n},\bar{m}}} \triangleq \mathbb{E}_{x} \big[ | x - \hat{x}_{\bar{n},\bar{m}}^{(i-1)} |^2 \big]= E_\mathrm{S} - |\hat{x}_{\bar{n},\bar{m}}^{(i-1)}|^2,
\label{eq:MSE_d_k}
\end{equation}
where $\mathbb{E}_{x}$ refers to expectation over all the possible symbols in the constellation $\mathcal{C}$. The \ac{GaBP} receiver for such a setup consists of three major stages, as described below.

\subsubsection{Soft Interference Cancellation} The objective of the \ac{sIC} stage at a given $i$-th iteration of the algorithm is to utilize the soft replicas $\hat{x}_{\bar{n},\bar{m}}^{(i-1)}$ from a previous iteration in order to calculate the data-centric \ac{sIC} signals $\tilde{y}_{x:\bar{n},\bar{m}}^{(i)}$. Exploiting expression~\eqref{General_I/O_arbitrary_elementwise}, the \ac{sIC} signals are given by
\begin{align}
\label{eq:d_soft_IC}
\tilde{y}_{x:\bar{n},\bar{m}}^{(i)} &= y_{\bar{n}} - \sum_{e \neq \bar{m}} h_{\bar{n},e} \hat{x}_{\bar{n},e}^{(i)}, \\
&= h_{\bar{n},\bar{m}} x_{\bar{m}} + \underbrace{\sum_{e \neq \bar{m}} h_{\bar{n},e}(x_e - \hat{x}_{\bar{n},e}^{(i)}) + \bar{w}_{\bar{n}}}_\text{interference + noise term}.
\end{align}
Leveraging the \ac{SGA}, the interference and noise terms in this expression can be approximated as Gaussian noise, such that the conditional \acp{PDF} of the \ac{sIC} signals become
\begin{equation}
\label{eq:cond_PDF_d}
\!\!p_{\tilde{\mathrm{y}}_{\mathrm{x}:\bar{n},\bar{m}}^{(i)} \mid \mathrm{x}_{\bar{m}}}(\tilde{y}_{x:\bar{n},\bar{m}}^{(i)}|x_{\bar{m}}) \propto \mathrm{exp}\bigg[ -\frac{|\tilde{y}_{x:\bar{n},\bar{m}}^{(i)}\! -\! h_{\bar{n},\bar{m}} x_{\bar{m}}|^2}{\tilde{\sigma}_{x:\bar{n},\bar{m}}^{2(i)}} \bigg]
\end{equation}
with their conditional variances expressed as follows:
\begin{equation}
\label{eq:soft_IC_var_d}
\tilde{\sigma}_{x:\bar{n},\bar{m}}^{2(i)} = \sum_{e \neq \bar{m}} \left|h_{\bar{n},e}\right|^2 \hat{\sigma}^{2(i)}_{x:{\bar{n},e}} + \sigma^2_w.
\end{equation}


\subsubsection{Belief Generation} 
In the belief generation stage of the algorithm, the \ac{SGA} is exploited under the assumptions that $\bar{N}$ is a sufficiently large number and that the individual estimation errors in $\hat{x}_{\bar{n},\bar{m}}^{(i-1)}$ are independent, in order to generate initial estimates (a.k.a. beliefs) for all the data symbols. As a consequence of the \ac{SGA} and with the conditional \acp{PDF} of~\eqref{eq:cond_PDF_d}, the following extrinsic \acp{PDF} are deduced:
\begin{equation}
\label{eq:extrinsic_PDF_d}
\prod_{e \neq \bar{n}} p_{\tilde{\mathrm{y}}_{\mathrm{x}:e,\bar{m}}^{(i)} \mid \mathrm{x}_{\bar{m}}}(\tilde{y}_{x:e,\bar{m}}^{(i)}|x_{\bar{m}}) \propto \mathrm{exp}\bigg[ - \frac{(x_{\bar{m}} - \bar{x}_{\bar{n},\bar{m}}^{(i)})^2}{\bar{\sigma}_{x:\bar{n},\bar{m}}^{2(i)}} \bigg],
\end{equation}
where the corresponding extrinsic means and variances are respectively defined as
\begin{equation}
\label{eq:extrinsic_mean_d}
\bar{x}_{\bar{n},\bar{m}}^{(i)} = \bar{\sigma}_{x:\bar{n},\bar{m}}^{(i)} \sum_{e \neq \bar{n}} \frac{h^*_{e,\bar{m}} \tilde{y}_{x:e,\bar{m}}^{(i)}}{ \tilde{\sigma}_{x:e,\bar{m}}^{2(i)}},
\end{equation}
\begin{equation}
\label{eq:extrinsic_var_d}
\bar{\sigma}_{x:\bar{n},\bar{m}}^{2(i)} = \bigg( \sum_{e \neq \bar{n}} \frac{|h_{e,\bar{m}}|^2}{\tilde{\sigma}_{x:e,\bar{m}}^{2(i)}} \bigg)^{\!\!\!-1}.
\end{equation}
with $h^*_{e,\bar{m}}$ denoting the complex conjugate of $h_{e,\bar{m}}$.

\begin{algorithm}[t]
    \caption{SIM Optimization and Data Detection}
    \label{alg:proposed_decoder}
    \setlength{\baselineskip}{11pt}
    \textbf{Input:} Receive signal vector $\vec{y}\in\mathbb{C}^{\bar{N}\times 1}$, complex channel matrix $\bar{\vec{H}} \in \mathbb{C}^{\bar{N}\times \bar{M}}$, number of \ac{GaBP} iterations $i_{\max}$, number of gradient descent iterations $i_{\mathrm{GD}}$, data constellation power $E_\mathrm{S}$, noise variance $\sigma^2_w$, and damping factor $\beta_x$. \\
    \textbf{Output:} $\hat{\vec{x}}$ 
    \vspace{-2ex} 
    \begin{algorithmic}[1]  
    \STATEx \hspace{-3.5ex}\hrulefill
    \STATEx \hspace{-3.5ex}\textbf{Initialization}
    \STATEx \hspace{-3.5ex} - Set iteration counter to $i=0$ and amplitudes $c_x = \sqrt{E_\mathrm{S}/2}$.
    \STATEx \hspace{-3.5ex} - Set initial data estimates to $\hat{x}_{\bar{n},\bar{m}}^{(0)} = 0$ and corresponding 
    variances to $\hat{\sigma}^{2(0)}_{x:{\bar{n},\bar{m}}} = E_\mathrm{S}$ $\forall \bar{n},\bar{m}$.
    \STATEx \hspace{-3.5ex}\hrulefill
    \STATEx \hspace{-3.5ex}\textbf{SIM Optimization via Steepest Ascent}
    \STATEx \hspace{-3.5ex}\textbf{for} $i=1$ to $i_{\mathrm{GD}}$ \textbf{do}: $\forall q, \tilde{q}, m, \tilde{m}$
    \STATE Compute the gradients from expressions \eqref{eq:final_grad_TXSIM} and \eqref{eq:final_grad_RXSIM}.
    \STATE Update normalization parameters from expression \eqref{eq:normalization_params}.
    \STATE Update the phase parameters via expression \eqref{eq:param_updates_GD}.
    
    \STATEx \hspace{-3.5ex}\textbf{end for}
    \STATEx \hspace{-3.5ex}\textbf{Data Detection via GaBP}
    \STATEx \hspace{-3.5ex}\textbf{for} $i=1$ to $i_\text{max}$ \textbf{do} $\forall \bar{n}, \bar{m}$
    \STATE Compute \ac{sIC} data signal $\tilde{y}_{x:{\bar{n},\bar{m}}}^{(i)}$ and its corresponding variance $\tilde{\sigma}^{2(i)}_{x:{\bar{n},\bar{m}}}$ from~\eqref{eq:d_soft_IC} and \eqref{eq:soft_IC_var_d}.
    \STATE Compute extrinsic data signal belief $\bar{x}_{\bar{n},\bar{m}}^{(i)}$ and its corresponding variance $\bar{\sigma}_{x:{\bar{n},\bar{m}}}^{2(i)}$ from~\eqref{eq:extrinsic_mean_d} and \eqref{eq:extrinsic_var_d}.
    \STATE Compute denoised and damped data signal estimate $\hat{x}_{\bar{n},\bar{m}}^{(i)}$ from~\eqref{eq:QPSK_denoiser} and \eqref{eq:d_damped}.
    \STATE Compute denoised and damped data signal variance $\hat{\sigma}_{x:{\bar{n},\bar{m}}}^{2(i)}$ from~\eqref{eq:MSE_d_k} and \eqref{eq:MSE_d_m_damped}.
    
    \STATEx \hspace{-3.5ex}\textbf{end for}
    \STATE Calculate $\hat{x}_{\bar{m}}$ $\forall \bar{m}$ (equivalently $\hat{\vec{x}}$) using~\eqref{eq:d_hat_final_est}. 
    
\end{algorithmic}
\end{algorithm}

\subsubsection{Soft Replica Generation} Finally, the soft replica generation stage consists of denoising the previously computed beliefs under a Bayes-optimal rule, in order to obtain the final estimates for the desired variables.
For \ac{QPSK} modulation\footnote{\ac{QPSK} is explicitly selected for simplicity, but \ac{wlg}, denoisers for other modulation schemes can be also designed \cite{TakahashiTCOM2019}.}, the Bayes-optimal denoiser is given by
\vspace{-1ex}
\begin{equation}
\hat{x}_{\bar{n},\bar{m}}^{(i)}\! =\! c_x \bigg(\! \text{tanh}\!\bigg[ 2c_d \frac{\Real{\bar{x}_{\bar{n},\bar{m}}^{(i)}}}{\bar{\sigma}_{x:{\bar{n},\bar{m}}}^{2(i)}} \bigg]\!\! +\! \jmath \text{tanh}\!\bigg[ 2c_d \frac{\Imag{\bar{x}_{\bar{n},\bar{m}}^{(i)}}}{\bar{\sigma}_{{x}:{\bar{n},\bar{m}}}^{2(i)}} \bigg]\!\bigg),\!\!
\label{eq:QPSK_denoiser}
\end{equation}
where $c_x \triangleq \sqrt{E_\mathrm{S}/2}$ denotes the magnitude of the real and imaginary parts of the explicitly chosen \ac{QPSK} symbols, with its corresponding variance updated as in~\eqref{eq:MSE_d_k}.

After obtaining $\hat{x}_{\bar{n},\bar{m}}^{(i)}$ as per expression~\eqref{eq:QPSK_denoiser}, the final outputs are computed by damping the results to prevent convergence to local minima due to incorrect hard-decision replicas \cite{Su_TSP_2015}.
Letting the damping factor be $0 < \beta_x < 1$, yields
\vspace{-1ex}
\begin{equation}
\label{eq:d_damped}
\hat{x}_{\bar{n},\bar{m}}^{(i)} = \beta_x \hat{x}_{\bar{n},\bar{m}}^{(i)} + (1 - \beta_x) \hat{x}_{\bar{n},\bar{m}}^{(i-1)},
\end{equation}
Similarly, the variances $\hat{\sigma}^{2(i)}_{x:{\bar{n},\bar{m}}}$ are first updated via~\eqref{eq:MSE_d_k} and then damped via
\vspace{-1ex}
\begin{equation}
\label{eq:MSE_d_m_damped}
\hat{\sigma}^{2(i)}_{x:{\bar{n},\bar{m}}} = \beta_x \hat{\sigma}_{x:{\bar{n},\bar{m}}}^{2(i)} + (1-\beta_x) \hat{\sigma}_{x:{\bar{n},\bar{m}}}^{2(i-1)},
\end{equation}
Finally, as result of the conflicting dimensions, the consensus update of the estimates can be obtained as follows:
\vspace{-1ex}
\begin{equation}
\label{eq:d_hat_final_est}
\hat{x}_{\bar{m}} = \left( \sum_{\bar{n}=1}^{\bar{N}} \frac{|h_{\bar{n},\bar{m}}|^2}{\tilde{\sigma}_{x:\bar{n},\bar{m}}^{2(i_\text{max})}} \right)^{\!\!\!-1} \! \! \left( \sum_{\bar{n}=1}^{\bar{N}} \frac{h^*_{\bar{n},\bar{m}} \tilde{y}_{x:\bar{n},\bar{m}}^{(i_\text{max})}}{ \tilde{\sigma}_{x:\bar{n},\bar{m}}^{2(i_\text{max})}} \right).
\end{equation}

The complete pseudocode for the \ac{SIM} parameterization and detection procedure here proposed is summarized in Algorithm~\ref{alg:proposed_decoder}.

\subsection{Performance Analysis}
In this section, simulation results for the \ac{BER} performances of \ac{OFDM}, \ac{OTFS} and \ac{AFDM} systems with \ac{QPSK} modulation in a \ac{SIM}-enabled \ac{DD} channel optimized via Algorithm~\ref{alg:proposed_decoder} are provided. For the sake of simplicity, uplink \ac{SIMO} and \ac{SISO} scenarios are considered where both the \ac{TX} and \ac{RX} are equipped with \ac{SIM}, and there are no \acp{RIS} in the environment.
%
All the other parameters are as in section \ref{subsec:comp_eff_channel}, with the exception that the number of paths here is set to a more realistic $P=5$ and the delay and Doppler shifts are randomly generated. In order to make sure that no power advantage other than the effect of the parametrized \ac{SIM} is provided, the complete channels are normalized such that $||\bar{\vec{H}}_\text{OFDM}||^2_F = ||\bar{\vec{H}}_\text{OTFS}||^2_F = ||\bar{\vec{H}}_\text{AFDM}||^2_F = ||\bar{\vec{H}}_\text{MIMO}||^2_F$ for all the cases.

\begin{figure}[H]
    \centering
    \captionsetup[subfloat]{labelfont=small,textfont=small}
    \subfloat[SIMO: $N = 256, \tilde{K} = \tilde{K}' = 16, P = 5, N_\mathrm{T} = 1, N_\mathrm{R} = 4, d_s = 1$.]{{\includegraphics[width=0.8\columnwidth]{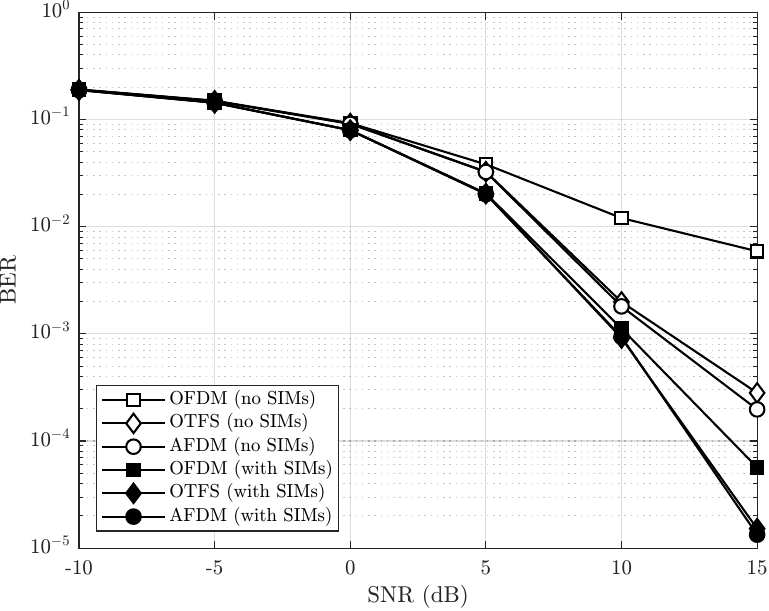}}}%
    \label{fig:BER_SIM_SIMO_4}
    \par\bigskip
    \vspace{-1ex}
    \subfloat[SIMO: $N = 256, \tilde{K} = \tilde{K}' = 16, P = 5, N_\mathrm{T} = 1, N_\mathrm{R} = 2, d_s = 1$.]{{\includegraphics[width=0.8\columnwidth]{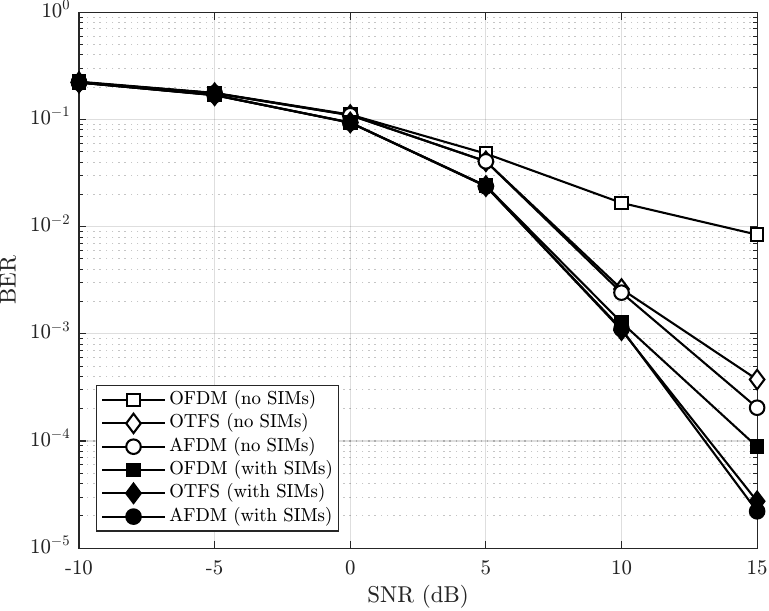}}}%
    \label{fig:BER_SIM_SIMO_2}
    \caption{\ac{BER} Performance of \ac{OFDM}, \ac{OTFS}, and \ac{AFDM} waveforms with \ac{QPSK} modulation in \ac{MPDD} channels with high-mobility, where \ac{SIM} has been placed at very close distances to both the \ac{TX} and the \ac{RX}.}
    \label{fig:BER_SIM}
\end{figure}

\begin{figure}[H]
    \centering
    {{\includegraphics[width=0.8\columnwidth]{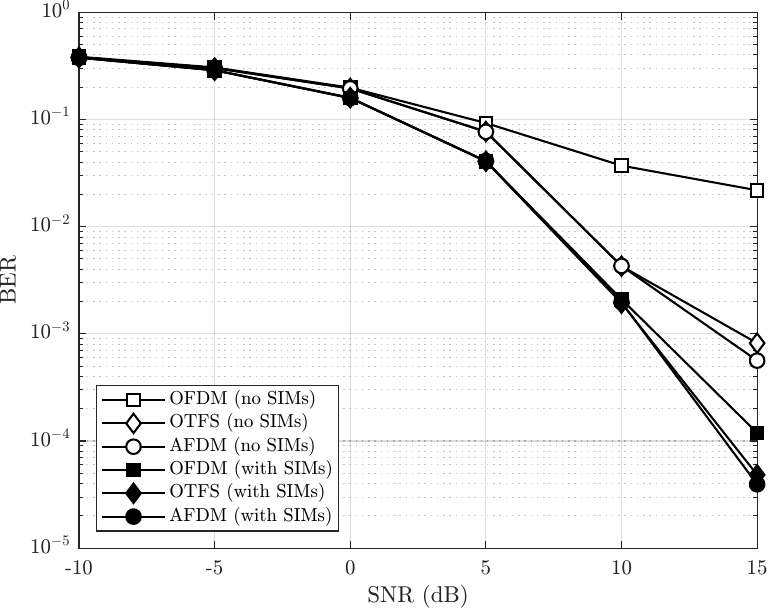}}}%
    \caption{Similar to Fig.~\ref{fig:BER_SIM} considering a \ac{SISO} setting.}
    \label{fig:BER_SIM_SISO}
\end{figure}

The results are shown in Fig.~\ref{fig:BER_SIM}, from which it can be seen that the \ac{SIM}-based systems significantly outperform those without \ac{SIM}. It can be observed that other than the obvious reduction of \ac{BER} resulting from the utilization of more receive antennas, the systems employing \ac{SIM} exhibit very similar behaviors (see Fig. \ref{fig:BER_SIM_SISO} for the \ac{SISO} results), which indicates that the technology indeed has the potential to substantially lower the gap between more sophisticated (multi-antenna) and less sophisticated (single-antenna) systems. In fact, it can also be seen that having \ac{SIM} significantly boosts the performance of conventional \ac{OFDM} compared to its typical performance in a \ac{DD} channel, bringing it close to that of far more sophisticated schemes, such as \ac{OTFS} and \ac{AFDM}.

\section{SIM Optimization and Radar Parameter Estimation}
\label{sec:SIM_Optimization_RPE}
The focus of this section is RPE performance (i.e., sensing performance), and investigating the role of \ac{SIM} for this functionality. 
To this end, the optimization problem presented in Section~\ref{SIM_Optimization} is reformulated to align towards a more sensing-centric approach, and solved via a greedy gradient ascent, and complemented by a \ac{PDA}-based estimator following \cite{KuranageTWC2024} to estimate the actual ranges and velocities of objects in the surrounding.

The system model considered is the same as Fig. \ref{fig:system_model}, with the only difference being that the goal is to estimate radar parameters at the receiver in a bistatic fashion, with full knowledge of the transmit signal via pilots or a front-haul. Note that a blind bistatic approach is also possible~\cite{RanasingheWCNC2024}. For simplicity, a \ac{SISO} system equipped with \ac{SIM} without \acp{RIS} in the ambient is considered henceforth. 

\subsection{SIM-Based Signal Enhancement}
\label{SIM_Optimization_for_sensing}

Similarly to Section~\ref{SIM_Optimization}, the matrices $\vec{\mathcal{Z}}$ and $\tilde{\vec{\mathcal{Z}}}$ must be parametrized to enhance the sensing performance by maximizing the power of the weakest path. To this end, define:
\begin{equation}
\label{eq:min_path_selection}
p_{\min} = \arg\underset{p}{\min} \; \Big|\Big| \vec{O}_p \Big|\Big|_F
\end{equation}
selecting the path with the minimum power, $\vec{O}_{p_{\min}}$, which can be maximized egoistically:
\begin{align}
\label{eq:path_optimization_sensing}
&\underset{\vec{\mathcal{Z}},\tilde{\vec{\mathcal{Z}}}}{\text{max}} \; \mathcal{O}_{\min}(\vec{\mathcal{Z}},\tilde{\vec{\mathcal{Z}}}) = \Big|\Big| \vec{O}_{p_{\min}}  \Big|\Big|_F^2\nonumber\\
&\;\;\;\text{\text{s}.\text{t}.}\;\vec{\Upsilon}_\mathrm{T}(\vec{\mathcal{Z}})\;\text{as in eq. \eqref{eq:transmit_SIM_full}},\;\vec{\Upsilon}_\mathrm{R}(\tilde{\vec{\mathcal{Z}}})\;\text{as in eq. \eqref{eq:receive_SIM_full}},\nonumber\\
&\;\;\;\quad\;\;\,\vec{\Psi}_q \; \text{and} \; \vec{\Delta}_{\tilde{q}}\;\text{as in eqs. \eqref{eq:diagona_shift_matrix_per_layer} and \eqref{eq:diagona_shift_matrix_per_layer_rx}, respectively},\nonumber\\
&\;\;\;\quad\;\;\,|\zeta^{q}_{m}| \leq \pi \; \forall (q,m), \text{and\;} \,|\tilde{\zeta}^{\tilde{q}}_{\tilde{m}}| \leq \pi \; \forall (\tilde{q},\tilde{m}).
\end{align}
In view of the non-convex unit modulus constraints, simple gradient-ascent technique is utilized in an iterative manner which tunes the phase shift parameters of the \ac{SIM} in a greedy fashion. Following Section~\ref{SIM_Optimization}, the gradient expressions can be derived as:
\begin{equation}
\label{eq:final_grad_TXSIM_sensing}
\nabla_{\vec{\zeta}_{q}} \mathcal{O}_{\min}(\vec{\mathcal{Z}},\tilde{\vec{\mathcal{Z}}}) = 2 \Im \Bigg\{ \sum_{n_t=1}^{N_T} \vec{\Psi}_{q}\herm \tilde{\vec{\Upsilon}}_{t:q,p_{\min},n_t}\herm \vec{o}_{p_{\min},n_t} \Bigg\},
\end{equation}
and
\begin{equation}
\label{eq:final_grad_RXSIM_sensing}
\nabla_{\tilde{\vec{\zeta}}_{\tilde{q}}} \mathcal{O}_{\min}(\vec{\mathcal{Z}},\tilde{\vec{\mathcal{Z}}}) = 2 \Im \Bigg\{ \sum_{n_t=1}^{N_T} \vec{\Delta}_{\tilde{q}}\herm \tilde{\vec{\Upsilon}}_{r:\tilde{q},p_{\min},n_t}\herm \vec{o}_{p_{\min},n_t} \Bigg\}.
\end{equation}

\subsection{PDA-Based Estimation}

Following \cite{KuranageTWC2024}, the \ac{PDA} procedure can be concisely described as follows.

\subsubsection{Problem Reformulation}
Leveraging the fact that the \ac{DD} channel can be characterized by a maximum normalized delay spread, $\text{max}(\ell_p)$, and a corresponding digital normalized Doppler spread, $\text{max}(f_p)$, satisfying the relations $\text{max}(\ell_p) << N$ and $\text{max}(f_p) << N$, for a finite number $P$ of paths, the channel impulse response in expression~\eqref{eq:SISO_DD_channel_model_DVIRF} can be rewritten as follows:
%
\begin{equation}
\label{eq:DD_channel_definition_frac_Dopp}
h^{\text{DD}}(\tau,\nu) = \sum_{\bar{k}=0}^{K_\tau - 1} \sum_{\bar{d}=0}^{D_\nu - 1} h_{\bar{k},\bar{d}} \delta(\tau - \tau_{\bar{k}}) \delta(\nu - \nu_{\bar{d}}),
\end{equation}
where $K_\tau$ and $D_\nu$ are numbers large enough to define a sufficiently fine grid discretizing the region determined by the maximum delay and Doppler values. Notice that, if the resolution of such a grid is sufficiently fine, the only non-zero terms inside the summation in~\eqref{eq:DD_channel_definition_frac_Dopp} are those where both $\tau_{\bar{k}}\approx \tau_p$ and $\nu_{\bar{d}} \approx \nu_p$, which in turn implies that estimating these radar parameters amounts to estimating the $P$ channel gains such that $h_{\bar{k},\bar{d}}\neq 0$. This suggests that the RPE problem can be formulated as a canonical sparse signal recovery problem, as follows.

First, rewriting the time-continuous model of~\eqref{eq:DD_channel_definition_frac_Dopp} in terms of the sampled equivalent in matrix form, yields the expression:
\begin{equation}
\label{eq:TD_channel_w_frac_Dopp_matrix}
\bar{\vec{H}}_\text{RPE} \triangleq \sum_{\bar{k}=0}^{K_\tau - 1} \sum_{\bar{d}=0}^{D_\nu - 1} \check{h}_{\bar{k},\bar{d}}^d  {\vec{G}_{\bar{k},\bar{d}}^\text{RPE}}.
\end{equation}
This model is identical to the equivalent channel implicitly defined in~\eqref{eq:H_DD_MIMO}, only with the summation rewritten over the \ac{DD} grid, as opposed to the channel paths, for a general waveform in a \ac{SISO} system. Then, the generalized receive signal model given in \eqref{General_I/O_arbitrary} can be re-expressed as follows: 
\begin{subequations}
\label{eq:sparse_RPE}
\vspace{-1ex}
\begin{equation}
\label{eq:modified_input_output_relation_for_estimation}
\vec{y} = \sum_{\bar{k}=0}^{K_\tau - 1} \sum_{\bar{d}=0}^{D_\nu - 1} \underbrace{{\vec{G}_{\bar{k},\bar{d}}^\text{RPE}} \vec{x}}_{\triangleq\vec{e}_{\bar{k},\bar{d}}\in \mathbb{C}^{N\times 1}} h_{\bar{k},\bar{d}} + \bar{\vec{w}} = \vec{E} \vec{h} + \bar{\vec{w}} \in \mathbb{C}^{N\times 1},
\end{equation}
where the vectors are defined as $\vec{e}_{\bar{k},\bar{d}}\triangleq{\vec{G}_{\bar{k},\bar{d}}^\text{RPE}} \vec{x}$ $\forall \bar{k},\bar{d}$, and explicitly identified the dictionary matrix $\vec{E} \in \mathbb{C}^{N\times K_\tau D_\nu}$ and the sparse channel vector $\vec{h} \in \mathbb{C}^{K_\tau D_\nu \times 1}$:
%
\begin{equation}
\label{eq:sparse_dict_matrix_def}
\vec{E}\! \triangleq\! [\vec{e}_{0,0}, \dots, \vec{e}_{0,D_\nu\! -\! 1}, \dots, \vec{e}_{K_\tau\! -\! 1,0}, \dots, \vec{e}_{K_\tau\! -\! 1,D_\nu \!-\! 1}],
\end{equation}
\begin{equation}
\label{eq:sparse_channel_vector}
\vec{h} \!\triangleq\! [h_{0,0}, \dots, h_{0,D_\nu\! -\! 1}, \dots, h_{K_\tau\! -\! 1,0}, \dots, h_{K_\tau\! -\! 1,D_\nu\! -\! 1}]\trans.
\end{equation}
\end{subequations}

In summary, the problem of estimating the set of radar parameters $\{\tau_p,\nu_p\}$ $\forall p$ reduces to estimating the sparse channel vector $\vec{h}$, with the delays and Doppler shifts obtained from the corresponding indices $(\bar{k},\bar{d})$ where $h_{\bar{k},\bar{d}}\neq 0$, given the received signal $\vec{y}$ and the dictionary matrix $\vec{E}$.
As a consequence of the approach, the assumption that $P$ is known is relaxed into the assumption that the channel paths are orthogonal in the delay-Doppler grid, such that $P$ can be unambiguously inferred from $\vec{h}$.
Finally, the method also implies the assumption that the transmit vector $\vec{x}$ is known at the radar receiver, which is in agreement with the considered bistatic setup.

\subsubsection{PDA-Based Radar Parameter Estimation}
In this section, a \ac{PDA}-based \ac{RPE} scheme designed under the assumption that the prior distributions of the elements $\hat{h}_{\check{m}}$\footnote{\Ac{wlg}, we set $\check{M} \triangleq K_\tau D_\nu$ for this section, with the corresponding elements $\check{m} = \{1,\dots,\check{M}\}$.} of the sparse channel vector estimates $\vec{\!\hat{\;h}}$ are of Bernoulli-Gaussian type is presented.
In other words, the unknown channel to be estimated at each $i$-th iteration of the proposed scheme is modelled as follows: 
%
\begin{equation}
\label{eq:h_m_estimate_dist}
\hat{h}_{\check{m}}^{(i)} \sim p_{\,\text{h}_{\check{m}}} (\text{h}_{\check{m}};\vec{\theta}^{(i)}),
\vspace{-2ex}
\end{equation}
with
%
\begin{equation}
\label{eq:h_m_pdf}
\!\!p_{\,\text{h}_{\check{m}}} (\text{h}_{\check{m}};\vec{\theta}^{(i)}) \triangleq\! (1\! -\! {\rho}^{(i)}) \delta(h_{\check{m}})\! + \!{\rho}^{(i)} \mathcal{CN}\big(h_{\check{m}};{\bar{h}}^{(i)},{\bar{\sigma}}^{(i)} \big),\!\!
\end{equation}
where $\vec{\theta}^{(i)} \triangleq [{\rho}^{(i)}, {\bar{h}}^{(i)},{\bar{\sigma}}^{(i)}]$ carries all three parameters of the distribution, namely, sparsity rate, mean, and variance. This parameter needs to be updated iteratively, and for this the \ac{EM} algorithm is considered. For later convenience, the soft replica (i.e., tentative estimate) of $h_{\check{m}}$ as $\big\{ \hat{h}_{{\check{m}}}^{(i)} \big\} $ can be defined, such that its \ac{MSE} can be expressed as follows:
%
\begin{equation}
\hat{\sigma}_{h:{\check{m}}}^{2(i)} \triangleq \mathbb{E}_{\text{h}_{\check{m}}} \big[  |h_{\check{m}} - \hat{h}_{{\check{m}}}^{(i)}|^2   \big].
\label{eq:MSE_h_PDA}
\end{equation}

Following steps similar to those in Section \ref{sec:SIM_Optimization_Detection}, the message passing algorithm is developed in the following manner.

\textbf{Soft Interference Cancellation:} The \ac{sIC} expression corresponding to an estimate of $h_{\check{m}}$, is given by
\vspace{-2ex}
\begin{equation}
\label{eq:Soft_IC_PDA}
\tilde{\vec{y}}_{h:{\check{m}}}^{(i)} = \vec{y} -\sum_{q\neq {\check{m}}} \vec{e}_{q} \hat{h}_{q}^{(i)}  = \vec{e}_{{\check{m}}}h_{\check{m}} + \overbrace{\sum_{q \neq {\check{m}}}^{K_\tau D_\nu}  (\vec{e}_{q} h_q - \vec{e}_{q} \hat{h}_{q}^{(i)}) + \tilde{\vec{w}}}^{\text{residual interference+noise component}}, 
\end{equation}
where $\vec{e}_{\check{m}}$ is the ${\check{m}}$-th column of the dictionary matrix $\vec{E}$. It follows from the \ac{CLT} that, under large-system conditions, the residual interference-plus-noise component can be approximated by a multivariate complex Gaussian variate.
In other words, the \ac{VGA} can be applied such that the conditional \ac{PDF} of the beliefs $\tilde{\vec{y}}_{h:{\check{m}}}^{(i)}$, given $h_{\check{m}}$, can be expressed as
\vspace{-0.5ex}
\begin{equation}
\label{eq:VGA_y_given_h}
\tilde{\vec{y}}_{h:{\check{m}}}^{(i)}\sim p_{\textbf{y} | \text{h}_{{\check{m}}}} (\textbf{y} | h_{{\check{m}}}) 
\propto \text{exp} \Big[ - \big( \textbf{y} - \vec{e}_{{\check{m}}} h_{{\check{m}}} \big)\herm \vec{\Sigma}^{-1(i)}_{{\check{m}}} \big( \textbf{y} - \vec{e}_{{\check{m}}} h_{{\check{m}}} \big) \Big],
\end{equation}
where $\textbf{y}$ is an auxiliary variable, and the conditional covariance matrix $\vec{\Sigma}_{\check{m}}^{(i)}$, with $\sigma_w^2$ being the noise power, is given by
\vspace{-1ex}
\begin{align}
\vec{\Sigma}_{\check{m}}^{(i)} & \triangleq \mathbb{E}_{\textbf{h,$\tilde{\vec{w}}$}|\hat{h}_{\check{m}} \neq h_{\check{m}}} \bigg[ \big( \tilde{\vec{y}}_{h:{\check{m}}}^{(i)} - \vec{e}_{{\check{m}}} h_{{\check{m}}} \big) \big( \tilde{\vec{y}}_{h:{\check{m}}}^{(i)} - \vec{e}_{{\check{m}}} h_{{\check{m}}} \big)\herm  \bigg]
\nonumber \\[-1ex]
& = \sum_{q \neq {\check{m}}}^{K_\tau D_\nu} \hat{\sigma}_{h:q}^{2(i)} \vec{e}_q \vec{e}_q\herm + \sigma_w^2 \vec{I}_N.
\label{eq:covariance_matrix_PDA}
\end{align}

\textbf{Belief Generation:} The beliefs associated with the estimate of the ${\check{m}}$-th channel entry $h_{\check{m}}$ can be obtained by combining the contributions of all \ac{sIC} beliefs $\tilde{\vec{y}}_{h:{\check{m}}}^{(i)}$, under the \ac{PDF}:
\begin{equation}
p_{\text{h} | h_{\check{m}}} (\text{h} | h_{\check{m}}) \propto \text{exp} \Big[ - \tfrac{|\text{h} - \tilde{h}_{\check{m}}^{(i)}|^2}{\tilde{\sigma}_{\tilde{h}:{\check{m}}}^{2(i)}} \Big],
\label{eq:ell_extrinsic_belief_PDA}
\end{equation}
which yields
\begin{subequations}
\label{eq:mean_and_var_extrinsic_belief_PDA}
\begin{equation}
\tilde{h}_{\check{m}}^{(i)} \triangleq \frac{1}{\eta_{\check{m}}^{(i)}} \vec{e}_{\check{m}}\herm \vec{\Sigma}^{-1(i)} \tilde{\vec{y}}_{h:{\check{m}}}^{(i)}\;\;\text{and}\;\;
\tilde{\sigma}_{\tilde{h}:{\check{m}}}^{2(i)} \triangleq \frac{1 - \eta_{\check{m}}^{(i)} \hat{\sigma}_{h:{\check{m}}}^{2(i)}}{\eta_{\check{m}}^{(i)}},
\end{equation}
where $\eta_{\check{m}}^{(i)}$ is a normalization factor defined as
\begin{equation}
\label{eq:eta_PDA}
\eta_{\check{m}}^{(i)} \triangleq \vec{e}_{\check{m}}\herm \vec{\Sigma}^{-1(i)} \vec{e}_{\check{m}},
\end{equation}
and the common conditional covariance matrix\footnote{The matrix inversion lemma is used in the derivation of~\eqref{eq:mean_and_var_extrinsic_belief_PDA} and by consequence \eqref{eq:eta_PDA}, such that the same inverse matrix $\vec{\Sigma}^{(i)}$ can be used instead of $\vec{\Sigma}_{\check{m}}^{(i)}$.}
is given by
\begin{equation}
\vec{\Sigma}^{(i)} \triangleq \sum_{{\check{m}}=1}^{K_\tau D_\nu} \hat{\sigma}_{h:{\check{m}}}^{2(i)} \vec{e}_{\check{m}} \vec{e}_{\check{m}}\herm + \sigma_w^2 \vec{I}_N.
\end{equation}
\end{subequations}

\textbf{Soft Replica Generation:} Under \ac{VGA}, as opposed to \ac{SGA} for \ac{GaBP}, the soft replicas of $h_{\check{m}}$ can be inferred from the conditional expectation given the extrinsic beliefs and the fact that the effective noise components in $\hat{h}_{\check{m}}^{(i)}, \forall {\check{m}}$ are uncorrelated, yielding:
\begin{equation}
p_{h_{{\check{m}}} | \text{h}} ( h_{{\check{m}}} | \text{h}; \vec{\theta}^{(i)}) = \frac{ p_{\text{h} | h_{{\check{m}}}} (\text{h} | h_{{\check{m}}};\tilde{h}_{\check{m}}^{(i)},\tilde{\sigma}_{\tilde{h}:{\check{m}}}^{2(i)}) \; p_{h_{{\check{m}}}}(h_{{\check{m}}}; \vec{\theta}^{(i)}) }{ \int_{h'_{{\check{m}}}} p_{\text{h} | h_{{\check{m}}}} (\text{h} | h'_{{\check{m}}};\tilde{h}_{\check{m}}^{(i)},\tilde{\sigma}_{\tilde{h}:{\check{m}}}^{2(i)}) \; p_{h_{{\check{m}}}} (h'_{{\check{m}}}; \vec{\theta}^{(i)}) }.
\label{eq:VGA_pdf_h_m_PDA}
\end{equation}
Next, leveraging the assumption that $h_{\check{m}}$ follows a Bernoulli-Gaussian distribution and using the Gaussian-\ac{PDF} multiplication rule \cite{Parker_TSP_2014}, \eqref{eq:VGA_pdf_h_m_PDA} can be rewritten as 
\begin{equation}
p_{h_{{\check{m}}} | \text{h}} ( h_{{\check{m}}} | \text{h}; \vec{\theta}^{(i)}) = (1 - \hat{\rho}_{{\check{m}}}^{(i)}) \delta(h_{\check{m}}) + \hat{\rho}_{{\check{m}}}^{(i)}\mathcal{CN}(h_{\check{m}};\hat{h}_{{\check{m}}}^{(i)},\hat{\sigma}_{h:{\check{m}}}^{2(i)}),
\label{eq:VGA_pdf_h_m_rewritten_BG_PDA}
\end{equation}
where
\begin{equation}
\label{eq:BG_update_sparsity_rate}
\hat{\rho}_{{\check{m}}}^{(i)} \triangleq \Bigg( \frac{1 - {\rho}^{(i)}}{{\rho}^{(i)}}  \frac{\tilde{\sigma}_{\tilde{h}:{\check{m}}}^{2(i)} + {\bar{\sigma}}^{(i)}}{\tilde{\sigma}_{\tilde{h}:{\check{m}}}^{2(i)}} \: e^{- \frac{|\tilde{h}_{{\check{m}}}^{(i)}|^2}{\tilde{\sigma}_{\tilde{h}:{\check{m}}}^{2(i)}} + \frac{|\tilde{h}_{{\check{m}}}^{(i)} - {\bar{h}}^{(i)}|^2}{\tilde{\sigma}_{\tilde{h}:{\check{m}}}^{2(i)} + {\bar{\sigma}}^{(i)}}} + 1 \Bigg)^{-1}
\end{equation}
with
\begin{equation}
\label{eq:BG_update_rules}
\hat{h}_{{\check{m}}}^{(i)} \triangleq \frac{{\bar{\sigma}}^{(i)} \tilde{h}_{{\check{m}}}^{(i)} + \tilde{\sigma}_{\tilde{h}:{\check{m}}}^{2(i)} {\bar{h}}^{(i)}}{\tilde{\sigma}_{\tilde{h}:{\check{m}}}^{2(i)} + {\bar{\sigma}}^{(i)}}\;\;\text{and}\;\;
\hat{\sigma}_{h:{\check{m}}}^{2(i)} \triangleq \frac{{\bar{\sigma}}^{(i)} \tilde{\sigma}_{\tilde{h}:{\check{m}}}^{2(i)}}{\tilde{\sigma}_{\tilde{h}:{\check{m}}}^{2(i)} + {\bar{\sigma}}^{(i)}}.
\end{equation}

Finally, adhering to the prior defined for the channel coefficient $h_p$, \ac{wlg}, a denoiser without the mean parameter, namely, ${\bar{h}}^{(i)} = 0$ in expression~\eqref{eq:h_m_pdf}, and consequently, in~\eqref{eq:BG_update_rules} can be chosen\footnote{For completeness, the \ac{EM} parameter update detailed in the consequent section will be derived for all the parameters.}.
From \eqref{eq:VGA_pdf_h_m_rewritten_BG_PDA}, the soft replica $\hat{h}_{{\check{m}}}^{(i)}$ and its \ac{MSE} $\hat{\sigma}_{h:{\check{m}}}^{2(i)}$ can be, in general, obtained from the conditional expectation as follows\footnote{Note the already incorporated damping procedure, as also done in Section \ref{sec:SIM_Optimization_Detection}, to prevent convergence to local minima due to incorrect hard-decision replicas.}:
\begin{subequations}
\label{eq:soft_rep_and_MSE_updates}
\begin{equation}
\label{eq:PDA_soft_rep_update}
\hat{h}_{\check{m}}^{(i)} = \tilde{\beta}_h \hat{\rho}_{{\check{m}}}^{(i)} \hat{h}_{{\check{m}}}^{(i)} + (1 - \tilde{\beta}_h) \hat{h}_{\check{m}}^{(i-1)},
\end{equation}
\begin{equation}
\label{eq:PDA_MSE_update}
\hat{\sigma}_{h:{\check{m}}}^{2(i)} = \tilde{\beta}_h \big[(1 - \hat{\rho}_{{\check{m}}}^{(i)}) \hat{\rho}_{{\check{m}}}^{(i)} |\hat{h}_{{\check{m}}}^{(i)}|^2 + \hat{\rho}_{{\check{m}}}^{(i)} \hat{\sigma}_{h:{\check{m}}}^{2(i)}\big] + (1-\tilde{\beta}_h) \big[ \hat{\sigma}_{h:{\check{m}}}^{2(i-1)} \big].
\end{equation}
\end{subequations}

\begin{algorithm}[t]
    \caption{SIM Optimization and RPE}
    \label{alg:proposed_estimator}
    \setlength{\baselineskip}{11pt}
    \textbf{Input:} Receive signal vector $\vec{y}$, dictionary matrix $\vec{E}$, number of paths $P$, average channel power per path $\sigma_h^2$, damping factor $\tilde{\beta}_h$, number of \ac{PDA} iterations $i_{\max}$, number of gradient descent iterations $i_{\mathrm{GD}}$, and noise variance $\sigma^2_w$. \\
    \textbf{Output:} Estimates $\hat{\tau}_p$ and $\hat{\nu}_p$ extracted from the non-zero indices of the sparse channel estimate vector $\!\hat{\;\vec{h}}$.
    \vspace{-2ex} 
    \begin{algorithmic}[1]  
    \STATEx \hspace{-3.5ex}\hrulefill
    \STATEx \hspace{-3.5ex}\textbf{Initialization}
    \STATEx \hspace{-3.5ex} - Set counter to $i = 0$, and initial distribution parameters ${\rho}^{(0)} = P/(K_\tau D_\nu)$ and ${\bar{\sigma}}^{(0)} = 1/P$.
    \STATEx \hspace{-3.5ex} - Set average channel power per path as $\sigma_h^2 = 1/(K_\tau D_\nu)$.
    \STATEx \hspace{-3.5ex} - Set initial estimates $\hat{h}_{{\check{m}}}^{(0)} = 0$ and ${\hat{\sigma}}_{h:{\check{m}}}^{(0)} = \sigma_h^2$ $\forall {\check{m}}$.
    \STATEx \hspace{-3.5ex}\hrulefill
    \STATEx \hspace{-3.5ex}\textbf{SIM Optimization via Greedy Steepest Ascent}
    \STATEx \hspace{-3.5ex}\textbf{for} $p=1$ to $P+1$
    \STATE Compute $p_{\min}$ using~\eqref{eq:min_path_selection}.
    \STATEx \textbf{for} $i=1$ to $i_{\mathrm{GD}}$ \textbf{do} $\forall q, \tilde{q}, m, \tilde{m}$
    \STATE \hspace{3.5ex} Compute the gradients using expressions~\eqref{eq:final_grad_TXSIM} and \eqref{eq:final_grad_RXSIM}.
    \STATE \hspace{3.5ex} Update normalization parameters from~\eqref{eq:normalization_params}.
    \STATE \hspace{3.5ex} Update the phase parameters via~\eqref{eq:param_updates_GD}.
    
    \STATEx \textbf{end for}
    \STATEx \hspace{-3.5ex}\textbf{end for}
    \STATEx \hspace{-3.5ex}\textbf{Radar Parameter Estimation via PDA}
    \STATEx \hspace{-3.5ex}\textbf{for} $i=1$ to $i_\text{max}$ \textbf{do} $\forall {\check{m}}$
    \STATE Compute soft signal vectors $\tilde{\vec{y}}_{h:{\check{m}}}^{(i)}$ from equation \eqref{eq:Soft_IC_PDA}.
    \STATE Compute soft extrinsic channel beliefs $\tilde{h}_{\check{m}}^{(i)}$ and their
    variances $\tilde{\sigma}_{\tilde{h}:{\check{m}}}^{2(i)}$ from \eqref{eq:mean_and_var_extrinsic_belief_PDA}.
    \STATE Compute denoised sparsity rates $\hat{\rho}_{\check{m}}^{(i)}$ from~\eqref{eq:BG_update_sparsity_rate}.
    \STATE Compute denoised channel estimates $\hat{h}_{{\check{m}}}^{(i)}$  and their variances $\hat{\sigma}_{h:{\check{m}}}^{2(i)}$ via \eqref{eq:BG_update_rules}.
    \STATE Compute damped channel estimates $\hat{h}_{{\check{m}}}^{(i)}$ and variances $\hat{\sigma}_{h:{\check{m}}}^{2(i)}$ using~\eqref{eq:soft_rep_and_MSE_updates}.
    \STATE Update distribution parameters ${\rho}^{(i)}$ and ${\bar{\sigma}}^{(i)}$ from \eqref{eq:Complete_set_update_BG}.
    \STATEx \hspace{-3.5ex}{\textbf{end}} {\textbf{for}} 
    \STATE Compute the estimates $\hat{\tau}_p$ and $\hat{\nu}_p$ corresponding to the indices ${\check{m}}$ of the non-zero entries of $\!\hat{\;\vec{h}}$ in accordance to expression~\eqref{eq:sparse_channel_vector}.
    
\end{algorithmic}
\end{algorithm}

\subsubsection{Parameter Update via EM}
In order to update the parameter set $\vec{\theta}$ of the Bernoulli-Gaussian distribution at each $i$-th iteration, the \ac{EM} algorithm is utilized.
In particular, \ac{EM} is employed as an iterative parameter tuning technique to obtain the parameter vector $\vec{\theta}$ that maximizes the likelihood function $p_{\vec{y} | \vec{\theta}} (\vec{y}|\vec{\theta})$, where $\vec{y}$ is defined in expression~\eqref{eq:modified_input_output_relation_for_estimation} with $\vec{\theta}$ given in~\eqref{eq:h_m_pdf}. To this end, following \cite{KuranageTWC2024}, one can obtain the update rules as follows:
\begin{subequations}
\begin{equation}
{\rho}^{(i)} = \frac{1}{K_\tau D_\nu} \sum_{{\check{m}}=1}^{K_\tau D_\nu} \hat{\rho}_{{\check{m}}}^{(i)},
\label{eq:Update_for_lambda}
\vspace{-0.75ex}
\end{equation}
\begin{equation}
{\bar{h}}^{(i)} = \frac{1}{{K_\tau D_\nu} {\rho}^{(i)}} \sum_{{\check{m}}=1}^{K_\tau D_\nu} \hat{\rho}_{{\check{m}}}^{(i)} \hat{h}_{{\check{m}}}^{(i)},
\label{eq:Update_for_mu}
\vspace{-0.75ex}
\end{equation}
\begin{equation}
{\bar{\sigma}}^{(i)} = \frac{1}{{K_\tau D_\nu} {\rho}^{(i)}} \sum_{{\check{m}}=1}^{K_\tau D_\nu} \hat{\rho}_{{\check{m}}}^{(i)} \bigg(  \big|  \hat{h}_{{\check{m}}}^{(i)} - {\bar{h}}^{(i)}  \big|^2 + \hat{\sigma}_{h:{\check{m}}}^{2(i)} \bigg).
\label{eq:Update_for_phi}
\vspace{-0.5ex}
\end{equation}
\label{eq:Complete_set_update_BG}
\end{subequations}

\begin{remark}[Assumptions]
    Since the complex path gains $\tilde{h}_p, \forall p$ in equation \eqref{eq:path_optimization_sensing} are generally unknown at the receiver for the \ac{SIM} optimization prior to the estimation procedure, these can be set to unity in order to initialize the receiver with equal path powers. 
    Additionally, the matrix $\vec{B}_p$ $\forall p$ is assumed to be known via a prior estimation; in general, one can use a \ac{DoA} estimation method as done in \cite{Ranasinghe_ICASSP_2024} to obtain a very accurate estimate of the \acp{AoA} and \acp{AoD}. 
\end{remark}

The complete pseudocode for the \ac{SIM} parameterization and sensing procedure proposed is summarized in Algorithm \ref{alg:proposed_estimator}.

\begin{figure}[t]
    \centering
    {{\includegraphics[width=0.8\columnwidth]{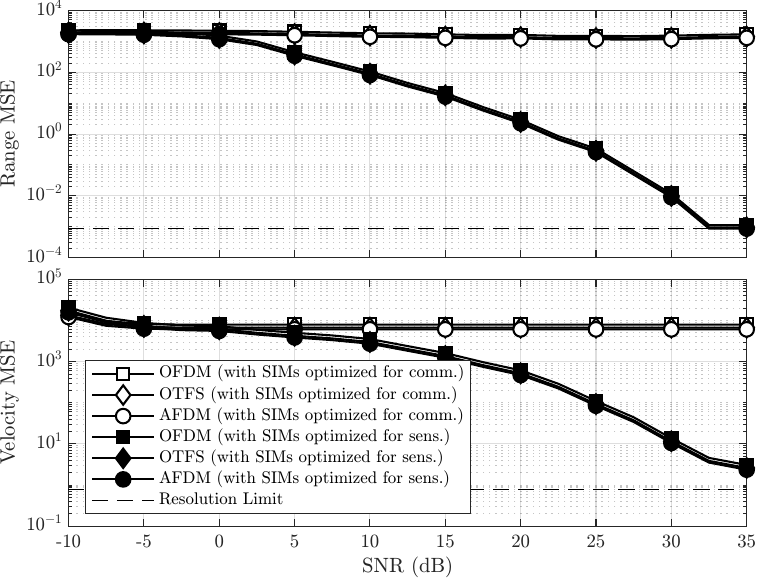}}}%
    \caption{\ac{MSE} performance of \ac{OFDM}, \ac{OTFS} and \ac{AFDM} waveforms with \ac{QPSK} modulation in \ac{MPDD} channels with high-mobility, with \ac{SIM} placed at very close distances to both the \ac{TX} and the \ac{RX} in a \ac{SISO} setting with $N = 144, \tilde{K} = \tilde{K}' = 12, P = 2, N_\mathrm{T} = 1, N_\mathrm{R} = 1, d_s = 1$.}
    \label{fig:MSE_SIM_SISO}
\end{figure}

\subsection{Sensing Performance Results}
\label{subsec:sensing_perf_analy}
Fig.~\ref{fig:MSE_SIM_SISO} illustrates the \ac{MSE} performance of the aforementioned method with \ac{OFDM}, \ac{OTFS}, and \ac{AFDM} signaling\footnote{It was shown in \cite{GaudioTWC2020,KuranageTWC2024} that the communication-centric \ac{ISAC} performance for \ac{OFDM}, \ac{OTFS}, and \ac{AFDM} are very similar, which holds true even with the \ac{SIM}.} with \ac{QPSK} modulation in a \ac{SIM}-enabled \ac{DD} channel, considering a \ac{SISO}
setting where both the \ac{TX} and \ac{RX} are equipped with \ac{SIM}, and there are no \acp{RIS} in the environment. It is assumed that the system operates with a carrier frequency of $28$ GHz ($i.e.$, at wavelength $\lambda = 10.7$ mm) with a bandwidth of $B = 20$ MHz in an environment with identical \ac{TX} and \ac{RX} \ac{SIM}, both with $Q = \tilde{Q} = 3$ layers of metasurfaces and $M = \tilde{M} = 100$ meta-atoms per layer, leading to $M_x = M_z = \tilde{M}_x = \tilde{M}_z = 10$. Finally, the sampling frequency is set to $F_\mathrm{S}=B$ and the number of symbols per frame to $N = 144$. For the \ac{MPDD} channel, the \ac{2D} and \ac{3D} elevation \acp{AoD}/\acp{AoA} are uniformly distributed in $[0,\pi]$, while the \ac{3D} azimuth \acp{AoD}/\acp{AoA} are in $[-\frac{\pi}{2},\frac{\pi}{2}]$. Within the channel, there are two targets ($i.e., P = 2$), with ranges $37.5$ m and $97.5$ m and velocities $-54$ m/s and $+54$ m/s, respectively. 

In Fig.~\ref{fig:MSE_SIM_SISO}, the resolution limit defines how fine the search grid is, i.e., it measures the estimation accuracy for the range and velocity if the estimate for $\!\hat{\;\vec{h}}$, obtained as the output of the proposed Algorithm~\ref{alg:proposed_estimator}, is perfect. As observed in the figure, there exists a large gain in performance when using the sensing-based optimization for the \ac{SIM}, as opposed to the communication-based optimization presented in the previous Section~\ref{sec:SIM_Optimization_Detection}.

\subsection{ISAC Performance Results}
\begin{figure}[!t]
    \centering
    {{\includegraphics[width=0.8\columnwidth]{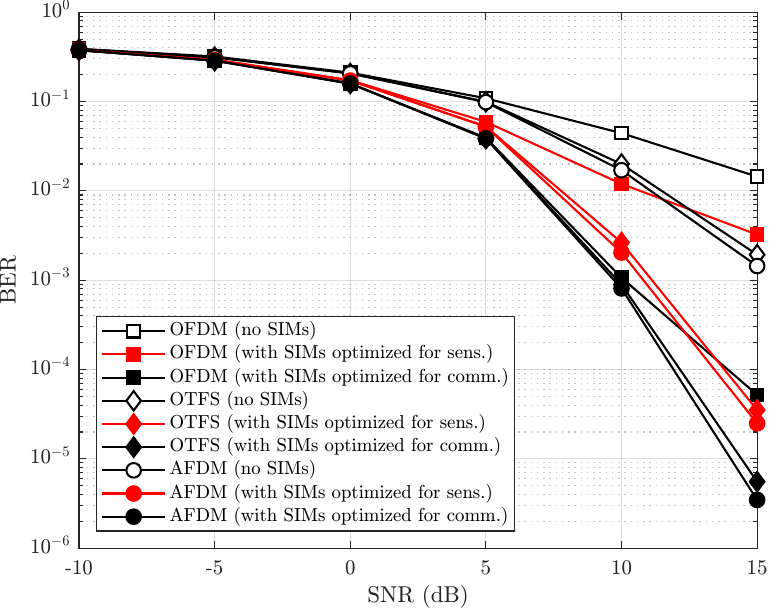}}}%
    \caption{\ac{BER} performance of \ac{OFDM}, \ac{OTFS} and \ac{AFDM} waveforms with \ac{QPSK} modulation in \ac{MPDD} channels with high-mobility, with \ac{SIM} placed at very close distances to both the \ac{TX} and the \ac{RX} in a \ac{SISO} setting with $N = 144, \tilde{K} = \tilde{K}' = 12, P = 2, N_\mathrm{T} = 1, N_\mathrm{R} = 1, d_s = 1$.}
    \label{fig:BER_SIM_SISO_ISAC}
\end{figure}

A natural question that occurs is whether there is a fundamental tradeoff between the performance of the communication and sensing functionalities due to the worse performance of the \ac{SIM} optimized for communications in a sensing setting. Under the same parameters as in the previous Section~\ref{subsec:sensing_perf_analy}, the communications performance of the \ac{SIM} optimized for sensing is also investigated. As seen from Fig.~\ref{fig:BER_SIM_SISO_ISAC}, while the communications performance decreases from the case where the \ac{SIM} is optimized for communications, there is still a significant gain over the case where no \ac{SIM} are used, highlighting the effectiveness of the \ac{SIM} for \ac{ISAC}, regardless of the waveform in use. 
However, it is noteworthy that the \ac{OFDM} waveform exhbits a smaller performance enhancement as opposed to \ac{OTFS} and \ac{AFDM} when using the sensing optimized \acp{SIM} for communications, hinting at the usual superiority of \ac{OTFS} and \ac{AFDM} waveforms in DD channels~\cite{KuranageTWC2024}. 

\vspace{-3ex}
\section{Conclusions}
\label{sec:Conclusion}
In this chapter, the \acl{MPDD} \ac{MIMO} channel model that integrates an arbitrary number of \acp{RIS} in the environment, together with \ac{SIM} at both the transmitter and receiver has been introduced, which is applicable to various \ac{ISAC}-enabling waveforms. By explicitly formulating the \ac{I/O} relationships for \ac{OFDM}, \ac{OTFS}, and \ac{AFDM}, it was demonstrated how the proposed \ac{MPDD} channel model could be effectively leveraged to optimize \ac{SIM}, enhancing the performance of these waveforms in \ac{DD} environments. The performance evaluation results indicated that \ac{SIM} significantly improve system performance, helping to bridge the gap between advanced waveform designs, such as \ac{OTFS} and \ac{AFDM}, with the conventional \ac{OFDM}.

\newpage

\end{document}